\documentclass[a4paper,11pt]{article}
\pdfoutput=1 % if your are submitting a pdflatex (i.e. if you have
\usepackage{jcappub} % for details on the use of th\citee package, please
\usepackage[T1]{fontenc} % if needed

\usepackage{amsmath}
\usepackage{natbib}
\usepackage{graphicx}
\usepackage{epsf}
\usepackage{color}
\usepackage{threeparttable}
\usepackage{comment}
\usepackage{epsfig}
\usepackage{xspace}
\usepackage[usenames,dvipsnames,svgnames]{xcolor}
\DeclareGraphicsExtensions{.jpg,.pdf,.png,.eps,.ps}

 \newcommand{\ltsima}{$\; \buildrel < \over \sim \;$}
 \newcommand{\ltsim}{\lower.5ex\hbox{\ltsima}}

 \newcommand{\planck}{\textit{Planck}} 
 \newcommand{\ninenine}{\ensuremath{95 \times 95}}
 \newcommand{\oneone}{\ensuremath{150 \times 150}}
 \newcommand{\twotwo}{\ensuremath{220 \times 220}}
 \newcommand{\nineone}{\ensuremath{95 \times 150}}
 \newcommand{\ninetwo}{\ensuremath{95 \times 220}}
 \newcommand{\onetwo}{\ensuremath{150 \times 220}}
 \newcommand{\chisq}{\ensuremath{\chi^2}}
 \newcommand{\chisqfull}{\ensuremath{236.3}}
 \newcommand{\chisqredfull}{\ensuremath{1.006}}
 \newcommand{\ptefull}{\ensuremath{0.464}}
 \newcommand{\ptefullpct}{\ensuremath{46\%}}
 \newcommand{\ytrue}{\ensuremath{y_\mathrm{true}}}
 \newcommand{\wtot}{\ensuremath{W_\mathrm{tot}}}

 \def\dl{$D_{\ell}$}

\title{\boldmath Consistency of cosmic microwave background temperature measurements in three frequency bands in the 2500-square-degree SPT-SZ survey}

\def\KICPChicago{1}
\def\AAUChicago{2}
\def\Oslo{3}
\def\Davis{4}
\def\FNAL{5}
\def\ArgonneHEP{6}
\def\PhysicsUChicago{7}
\def\EFIChicago{8}
\def\SLAC{9}
\def\McGill{10}
\def\Caltech{11}
\def\Berkeley{12}
\def\Cifar{13}
\def\Colorado{14}
\def\ESO{15}
\def\Colphys{16}
\def\Illast{17}
\def\Illphys{18}
\def\UChicago{19}
\def\LBNL{20}
\def\Arizona{21}
\def\Michigan{22}
\def\Munich{23}
\def\ExcellenceCluster{24}
\def\MPE{25}
\def\KIPAC{26}
\def\Stanford{27}
\def\Minnesota{28}
\def\Melbourne{29}
\def\CaseWestern{30}
\def\ArtInstChicago{31}
\def\JPL{32}
\def\CfA{33}
\def\Dunlap{34}
\def\UToronto{35}

\author[\KICPChicago,\AAUChicago,\Oslo]{L.~M.~Mocanu,}
\author[\KICPChicago,\AAUChicago,a]{T.~M.~Crawford,\note[a]{Corresponding author.}}
\author[\Davis]{K.~Aylor,}
\author[\FNAL,\KICPChicago,\AAUChicago]{B.~A.~Benson,}
\author[\ArgonneHEP,\KICPChicago]{L.~E.~Bleem,}
\author[\KICPChicago,\ArgonneHEP,\PhysicsUChicago,\AAUChicago,\EFIChicago]{J.~E.~Carlstrom,}
\author[\ArgonneHEP,\KICPChicago,\AAUChicago]{C.~L.~Chang,}
\author[\SLAC]{H-M.~Cho,}
\author[\McGill]{R.~Chown,}
\author[\KICPChicago,\AAUChicago,\Caltech]{A.~T.~Crites,}
\author[\McGill,\Berkeley]{T.~de~Haan,}
\author[\McGill,\Cifar]{M.~A.~Dobbs,}
\author[\Colorado]{W.~B.~Everett,}
\author[\Berkeley,\ESO]{E.~M.~George,}
\author[\Colorado,\Colphys]{N.~W.~Halverson,}
\author[\Berkeley]{N.~L.~Harrington,}
\author[\ArgonneHEP,\KICPChicago]{J.~W.~Henning,}
\author[\McGill,\Cifar,\Illast,\Illphys]{G.~P.~Holder,}
\author[\Berkeley]{W.~L.~Holzapfel,}
\author[\KICPChicago,\AAUChicago]{Z.~Hou,}
\author[\UChicago]{J.~D.~Hrubes,}
\author[\Davis]{L.~Knox,}
\author[\Berkeley,\LBNL]{A.~T.~Lee,}
\author[\UChicago]{D.~Luong-Van,}
\author[\Arizona]{D.~P.~Marrone,}
\author[\Michigan]{J.~J.~McMahon,}
\author[\KICPChicago,\AAUChicago,\EFIChicago,\PhysicsUChicago]{S.~S.~Meyer,}
\author[\Davis]{M.~Millea,}
\author[\Munich,\ExcellenceCluster,\MPE]{J.~J.~Mohr,}
\author[\AAUChicago]{T.~Natoli,}
\author[\McGill,\KIPAC,\Stanford]{Y.~Omori,}
\author[\KICPChicago,\AAUChicago]{S.~Padin,}
\author[\Minnesota]{C.~Pryke,}
\author[\Melbourne]{C.~L.~Reichardt,}
\author[\CaseWestern]{J.~E.~Ruhl,}
\author[\CaseWestern,\Colorado]{J.~T.~Sayre,}
\author[\KICPChicago,\EFIChicago,\ArtInstChicago]{K.~K.~Schaffer,}
\author[\Berkeley,\KICPChicago,\AAUChicago]{E.~Shirokoff,} 
\author[\CaseWestern,\JPL]{Z.~Staniszewski,}
\author[\CfA]{A.~A.~Stark,}
\author[\KIPAC,\Stanford]{K.~T.~Story,}
\author[\Dunlap,\UToronto]{K.~Vanderlinde,}
\author[\Illast,\Illphys]{J.~D.~Vieira,}
\author[\KICPChicago,\AAUChicago]{R.~Williamson,}
\author[\KICPChicago]{W.~L.~K.~Wu}

\affiliation[\KICPChicago]{Kavli Institute for Cosmological Physics, University of Chicago, Chicago, IL, USA 60637}
\affiliation[\AAUChicago]{Department of Astronomy and Astrophysics, University of Chicago, Chicago, IL, USA 60637}
\affiliation[\Oslo]{Institute of Theoretical Astrophysics, University of Oslo, P.O. Box 1029 Blindern, NO-0315 Oslo, Norway}
\affiliation[\Davis]{Department of Physics, University of California, Davis, CA, USA 95616}
\affiliation[\FNAL]{Fermi National Accelerator Laboratory, MS209, P.O. Box 500, Batavia, IL 60510}
\affiliation[\ArgonneHEP]{High Energy Physics Division, Argonne National Laboratory, Argonne, IL, USA 60439}
\affiliation[\PhysicsUChicago]{Department of Physics, University of Chicago, Chicago, IL, USA 60637}
\affiliation[\EFIChicago]{Enrico Fermi Institute, University of Chicago, Chicago, IL, USA 60637}
\affiliation[\SLAC]{SLAC National Accelerator Laboratory, 2575 Sand Hill Road, Menlo Park, CA 94025}
\affiliation[\McGill]{Department of Physics and McGill Space Institute, McGill University, Montreal, Quebec H3A 2T8, Canada}
\affiliation[\Caltech]{California Institute of Technology, Pasadena, CA, USA 91125}
\affiliation[\Berkeley]{Department of Physics, University of California, Berkeley, CA, USA 94720}
\affiliation[\Cifar]{Canadian Institute for Advanced Research, CIFAR Program in Cosmology and Gravity, Toronto, ON, M5G 1Z8, Canada}
\affiliation[\Colorado]{Center for Astrophysics and Space Astronomy, Department of Astrophysical and Planetary Sciences, University of Colorado, Boulder, CO, 80309}
\affiliation[\ESO]{European Southern Observatory, Karl-Schwarzschild-Stra{\ss}e 2, 85748 Garching, Germany}
\affiliation[\Colphys]{Department of Physics, University of Colorado, Boulder, CO, 80309}
\affiliation[\Illast]{Astronomy Department, University of Illinois at Urbana-Champaign, 1002 W. Green Street, Urbana, IL 61801, USA}
\affiliation[\Illphys]{Department of Physics, University of Illinois Urbana-Champaign, 1110 W. Green Street, Urbana, IL 61801, USA}
\affiliation[\UChicago]{University of Chicago, Chicago, IL, USA 60637}
\affiliation[\LBNL]{Physics Division, Lawrence Berkeley National Laboratory, Berkeley, CA, USA 94720}
\affiliation[\Arizona]{Steward Observatory, University of Arizona, 933 North Cherry Avenue, Tucson, AZ 85721}
\affiliation[\Michigan]{Department of Physics, University of Michigan, Ann  Arbor, MI, USA 48109}
\affiliation[\Munich]{Faculty of Physics, Ludwig-Maximilians-Universit\"{a}t, 81679 M\"{u}nchen, Germany}
\affiliation[\ExcellenceCluster]{Excellence Cluster Universe, 85748 Garching, Germany}
\affiliation[\MPE]{Max-Planck-Institut f\"{u}r extraterrestrische Physik, 85748 Garching, Germany}
\affiliation[\Stanford]{Dept. of Physics, Stanford University, 382 Via Pueblo Mall, Stanford, CA 94305}
\affiliation[\KIPAC]{Kavli Institute for Particle Astrophysics and Cosmology, Stanford University, 452 Lomita Mall, Stanford, CA 94305}
\affiliation[\Minnesota]{Department of Physics, University of Minnesota, Minneapolis, MN, USA 55455}
\affiliation[\Melbourne]{School of Physics, University of Melbourne, Parkville, VIC 3010, Australia}
\affiliation[\CaseWestern]{Physics Department, Center for Education and Research in Cosmology and Astrophysics, Case Western Reserve University,Cleveland, OH, USA 44106}
\affiliation[\ArtInstChicago]{Liberal Arts Department, School of the Art Institute of Chicago, Chicago, IL, USA 60603}
\affiliation[\JPL]{Jet Propulsion Laboratory, California Institute of Technology, Pasadena, CA 91109, USA}
\affiliation[\CfA]{Harvard-Smithsonian Center for Astrophysics, Cambridge, MA, USA 02138}
\affiliation[\Dunlap]{Dunlap Institute for Astronomy \& Astrophysics, University of Toronto, 50 St George St, Toronto, ON, M5S 3H4, Canada}
\affiliation[\UToronto]{Department of Astronomy \& Astrophysics, University of Toronto, 50 St George St, Toronto, ON, M5S 3H4, Canada}

\emailAdd{tcrawfor@kicp.uchicago.edu}

\abstract{We present an internal consistency test of South Pole Telescope (SPT) measurements of the cosmic microwave background (CMB) temperature anisotropy using three-band data from the SPT-SZ survey. These measurements are made from observations of $\sim$2500 deg$^2$ of sky in three frequency bands centered at 95, 150, and 220 GHz. We combine the information from these three bands into six semi-independent estimates of the CMB power spectrum (three single-frequency power spectra and three cross-frequency spectra) over the multipole range $650<\ell<3000$. We subtract an estimate of foreground power from each power spectrum and evaluate the consistency among the resulting CMB-only spectra. 
We determine
that the six foreground-cleaned power spectra are consistent with the 
null hypothesis, in which the six cleaned spectra contain only CMB power and noise.  A fit of
the data to this model results in a \chisq\ value of \chisqfull\ for 235 degrees of freedom, and the probability to exceed this \chisq\ value is \ptefullpct.
}

\begin{document}
\maketitle
\flushbottom

%%%%%%%%%%%%%%%%%%%%%%%%%%%%%%%%%%%%
% INTRODUCTION
%%%%%%%%%%%%%%%%%%%%%%%%%%%%%%%%%%%%

\section{Introduction}
\label{sec:intro}

Measurements of the temperature anisotropy in the cosmic microwave background (CMB) have
played a key role in establishing our current understanding of the Universe. Since the first detection of 
anisotropy in data from the Cosmic Background Explorer Differential Microwave Radiometer experiment
\cite{smoot92}, the power of CMB temperature anisotropy measurements to constrain cosmology
has progressed steadily, culminating in measurements from the \planck\ satellite \cite{planck15-13}.
The \planck\ measurements achieve percent-level constraints on five of the six parameters of the
so-called Lambda Cold Dark Matter model.

At this level of statistical precision, certain mild tensions have arisen among and within 
CMB datasets, and between
CMB data and other cosmological measurements. 
One of these is a  mild tension between the \planck\ best-fit
cosmological parameters and those estimated from a combination of 
data from the South Pole Telescope (SPT) and the \textit{Wilkinson Microwave Anisotropy Probe}
(\textit{WMAP}) in \cite[][hereafter S13]{story13}. 
These tensions have been noted 
in several publications including \cite{planck13-16}. The cosmological constraints in S13 
are tighter than any CMB constraints in the 
literature other than the \planck\ constraints, and it is important to understand whether disagreements in 
the two datasets arise from slightly unusual statistical fluctuations, from untreated systematic effects
in either dataset, or from a breakdown in our cosmological model. References \cite{hou18}
and \cite{aylor17} compared data from SPT and \planck\ in map, power spectrum, and cosmological
parameter space in an attempt to distinguish among these possibilities and found no evidence 
that the parameter differences are due to unmodeled systematic errors in either data set.

In this work, we specifically target the possibility of unmodeled systematics in the SPT data by
comparing data from three frequency bands in the 2500-square-degree SPT-SZ survey.
We refer to the three bands as 95~GHz, 150~GHz, and 220~GHz, indicating their rough center frequencies.
The 150~GHz SPT-SZ data provided the bulk of the statistical weight to 
the S13 cosmological constraints, so this comparison
is directly relevant to the investigation of cosmological parameter differences between S13 and \planck.
Additive and multiplicative systematic errors typical of CMB datasets are unlikely to be 
100\% correlated between observing bands, so a comparison of CMB temperature measurements in
different bands can be a powerful tool for identifying such errors. Typical additive 
systematics include incomplete or inaccurate foreground modeling and subtraction, inaccurate 
noise modeling, and ground pickup (for ground-based telescopes). 
The most important multiplicative systematics for CMB temperature measurements are 
inaccurate modeling of the telescope beam and absolute response. All of these are likely
to vary with observing frequency, depending on how beams, noise, and calibration are characterized.

For the particular case of SPT-SZ data, the most difficult residual systematics to rule out as potential 
causes for tension with \planck\ are estimates of foregrounds and beams. Systematic errors in both of these
quantities are expected to have very different amplitudes in the three observing bands, because of the spectral
dependence of millimeter-wave emission mechanisms and the different beam sizes in the three 
SPT-SZ bands. Thus, we would expect significant systematics in either of these quantities to lead to 
statistical disagreement among the foreground-cleaned CMB temperature estimates in the three bands.

We estimate the consistency among CMB measurements in the three SPT-SZ bands by 
creating six angular power spectra, including the three single-frequency power spectra and three 
cross-frequency spectra (\nineone, \ninetwo, and \onetwo). We subtract an estimate of foregrounds from each
spectrum, and we fit the six foreground-cleaned spectra to a simple model in which the CMB is the only
source of signal remaining in the spectra. The \chisq\ of this fit and the
probability to exceed (PTE) that value of \chisq\ are
our metrics for internal consistency of this dataset. 
We note that a version of this test has already been implicitly performed in the
cosmological analysis of \cite[][hereafter G15]{george15}. In that work, the same three-band data
were used to estimate the same six power spectra, and the spectra were fit to a combination CMB and 
foreground model in the multipole range $2000 <  \ell < 11000$. The CMB part of the model nominally 
came from a six-parameter $\Lambda$CDM model, but the fit was performed jointly with \planck\ and
baryon acoustic oscillation data, which effectively fixed the cosmological parameters and hence the
predicted CMB power spectrum. The result of that fit was a \chisq\
of 88 for 80 degrees of freedom and a PTE of 21.5\%. 
The test in this paper differs from G15 in multipole
range and model assumptions. In this work we examine the multipole range used for cosmological 
constraints in S13, namely $650 < \ell < 3000$, and we use only data above $\ell = 3000$ to 
construct the foreground model.
Assuming nothing about the underlying 
cosmological model, we test whether the six foreground-cleaned power spectra yield consistent estimates of the power in each $\ell$ bin.
%Assuming an acceptable level of consistency is 
%demonstrated, we declare the output of the fit to be our best CMB-only power spectrum and fit this
%power spectrum to a cosmological model, comparing to other recent SPT results and \planck.

This paper is organized as follows. 
We describe the SPT telescope and SPT-SZ camera, the 
SPT-SZ survey observations, and the data analysis up to the map level in Section~\ref{sec:obs}.
We describe our power spectrum pipeline in Section~\ref{sec:ps},
our construction of the $\ell$-space covariance matrix in Section~\ref{sec:cov}, and
a series of null tests we perform on the data in Section~\ref{sec:systematicTests}.
We describe our treatment of foregrounds in Section~\ref{sec:foregroundTreatment}.
The method we use to test for inter-band consistency is presented in Section~\ref{sec:consistall}.
We present the individual power spectra and the results of the consistency test in
Section~\ref{sec:results}, and we conclude in Section~\ref{sec:conclusion}.

\section{Instrument, observations, and data reduction}
\label{sec:obs}

The SPT is a 10-meter telescope located at the Amundsen-Scott South Pole station in Antarctica. 
The first camera on the SPT, the SPT-SZ camera, was a 960-pixel transition-edge-sensor bolometer
receiver configured to observe in three frequency bands centered at roughly 95, 150, and 220~GHz.
For more information on the telescope and camera, see \cite{padin08} and \cite{carlstrom11}.
From 2008-2011, the SPT-SZ camera was used to conduct a survey
covering a $\sim$2500 deg$^2$ region of sky between declinations of -40$^\circ$ and -65$^\circ$ and right ascensions of 20h and 7h. The SPT-SZ survey footprint is shown in, e.g., Figure 1 of S13 and Figure 1 of 
\cite{hou18}.
% Here we present the first power spectrum measurement that uses data from the complete SPT-SZ survey.
% We use data from 2540 deg$^2$ of sky in this analysis.
The SPT-SZ survey area was observed in 19 contiguous sub-patches, or fields, ranging in area from 
$\sim$70 to $\sim$225 deg$^2$. 
%The locations and sizes of the 19 fields are given in Table~\ref{tab:fields}.
%The effective areas in Table~\ref{tab:fields} differ slightly from those in Table 1 of S13 because we are using
%all three frequency bands in this work (as opposed to 150~GHz only in S13), and the coverage patterns
%are slightly different among the three bands. 

The observations and data reduction methods used in this work are very similar to those 
described in \cite[][hereafter K11]{keisler11} and S13, and we refer the reader to those works for detailed descriptions of the
observations and analysis. In the following, we give a brief overview of the most important
features of the observations and data reduction.

The SPT-SZ fields are observed using a raster-scan pattern, in which the telescope is scanned 
back and forth across the field at constant elevation. After each right/left scan pair, a small step in 
elevation is taken, and the process is repeated until the full field is covered. Though the extent in elevation
and the size of the elevation step varies field to field, a typical single-field observation takes roughly
two hours. Each field is observed an average of $\sim$200 times. Two SPT-SZ fields were also observed 
with elevation scans at constant azimuth, but only the azimuth-scan data are included in this analysis.
The SPT-SZ detector array was upgraded between the 2008 and 2009 seasons; in 2008 the 95~GHz
detectors did not produce science-quality data. The two fields that were observed in 2008 were 
reobserved later to gain 95~GHz sensitivity on these fields, but for ease of building on the S13
analysis we only use 2008 data on these fields in this work.

The raw time-ordered data (TOD) from the SPT-SZ survey are converted to maps of the sky 
using a simple bin-and-average process. Maps are made individually from each single observation
of a field, using data from all detectors of a given observing frequency. For a given map pixel, the TOD samples 
from any detector of a given frequency pointed in that direction in the sky are combined using 
inverse-noise-weighted averaging. Each detector is assigned a single weight value in an observation, 
based on a combination of noise in a certain range of temporal frequencies and the detector response
to celestial sources. This mapmaking procedure returns the minimum-variance result in the limit of
detector noise that is ``white'' (uncorrelated between time samples) and uncorrelated between detectors.
To reduce correlated noise among time samples for a given detector and among detectors, and to 
reduce aliasing of high-frequency noise when we bin into pixels, we filter the TOD before mapmaking.
The TOD from each detector are high-pass filtered by projecting out across a single scan a set of low-order polynomials and sines and cosines. The high-pass filter cutoff corresponds to roughly $\ell=300$ in the scan direction.
The anti-aliasing filter is a Fourier-domain low-pass filter with
a cutoff at a temporal frequency corresponding to roughly $\ell=6600$ in the scan direction.
Finally, 
%Three spatial modes---a two-dimensional plane with a constant offset---are fit to the data 
the average over all detectors in each of the six detector modules is subtracted from every detector 
at every time sample.
%containing 
%160 of the 960 SPT-SZ detectors, all at a single frequency) are subtracted from the data from each
%detector in the wedge 
%at each time sample, and the best-fit model is subtracted from every detector in the module.
In the pseudo-$C_\ell$ power spectrum pipeline described in the next section, we account for 
the effects of this filtering through the filter transfer function. 
%, as well as our knowledge of instrument beam and calibration. 

We estimate the filter transfer function using simulated observations. We create simulated skies
with lensed CMB fluctuations (with an underlying power spectrum from the best-fit 
\planck\ 2015 {\tt TT,TE,EE+lowP+lensing}
cosmological model \cite{planck15-13})
and foregrounds, and we mock-observe these skies using the real 
detector pointing, weights, and filtering. We create maps from these simulated data using the same
procedure as for real data. We calculate the power spectrum of these mock maps and compare it 
to the known input power spectrum. We define the filter transfer function $F_\ell$ following Eqn.~18 in \citep{hivon02}.
 We create a separate filter transfer function for each observing band.
%Typical filter transfer functions for SPT maps are shown in, e.g., \cite{schaffer11}.

The SPT-SZ beams are measured with a combination of planet observations and sources in the
SPT-SZ fields. The main lobes of the beams are measured using bright sources in the fields, 
the sidelobes are measured using observations of Jupiter, and these two measurements are connected using
observations of Venus. Individual beam profiles are estimated for each frequency band and each
survey year (because the receiver was modified between every observing season). We approximate
the beams as azimuthally symmetric and create a one-dimensional $\ell$-space beam function 
$B_\ell$ for each frequency band and year.

The relative calibration between detectors and between individual observations is accomplished using
a combination of observations of the Galactic HII region RCW38 and the response of detectors to 
a blackbody source mounted behind the secondary mirror. An approximate calibration for each frequency band is obtained using the 
known brightness of RCW38 in these bands. 
%The final absolute calibration is obtained by comparing the fully coadded 150~GHz field maps to 
%\planck\ and comparing the 95 and 220~GHz coadded maps to the 150~GHz maps. 
The final absolute calibration for each band is obtained either by comparing the fully coadded 
maps to \planck\ or by comparing power spectra of those maps to \planck\ power spectra.
%For 150~GHz, 
%we use the best-fit calibration from the \planck\ comparison in \cite{hou18}. For 95 and 220~GHz, 
%we use an absolute calibration obtained by comparing G15 power spectra with
%full-sky \planck\ 2015 power spectra.
At 150 GHz, we use the absolute calibration from \cite{hou18}, which was obtained using a cross-spectrum analysis between the SPT and \planck\ maps.  At 95 and 220~GHz, we use 
an absolute calibration obtained by comparing SPT power spectra from G15 with
full-sky \planck\ 2015 power spectra.
%, which was derived from a comparison of SPT power
%spectra to full-sky \planck\ power spectra. 
The fractional uncertainty on the absolute calibration is 2.1\%, 0.3\%, and 4.5\% in power
in the 95, 150, and 220~GHz bands, respectively. We use these uncertainties 
to derive the calibration contribution to the covariance described in Section~\ref{sec:cov}.

We note that we could in principle
achieve tighter priors on the 95 and 220~GHz calibrations by doing a map-based comparison
to \planck\ similar to that done in \cite{hou18} for 150~GHz. We can achieve higher precision, however,
by transferring the 150~GHz calibration to 95 and 220~GHz using our own data. If we were to 
perform the fit in this paper with the 95 and 220~GHz calibrations as free parameters
(instead of folding the calibration uncertainty into the bandpower covariance matrix), the 
posteriors on those parameters would be our best possible calibration in those bands. We have
in fact performed such a fit, the posteriors on the best-fit 95 and 220~GHz calibrations relative
to the priors are 
$1.016 \pm 0.004$ and $1.046 \pm 0.009$ (in power), respectively, and we will use these calibrations in future analyses 
of 95 and 220~GHz SPT-SZ data.

%%%%%%%%%%%%%%%%%%%%%%%%%%%%%%%%%%%%
% Power Spectrum
%%%%%%%%%%%%%%%%%%%%%%%%%%%%%%%%%%%%
\section{Power Spectrum Estimation}
\label{sec:ps}

In this section, we describe the pipeline used to compute the angular power spectrum from the maps. 
This analysis follows the methods used in \cite{lueker10}, \cite{shirokoff11}, K11, S13, and G15,
and we refer the reader to those works for more detail.
We adopt the flat-sky approximation, in which angular wavenumber $k$ is equivalent to multipole number $\ell$,
and spherical harmonic transforms are replaced by Fourier transforms. 
The six power spectra are calculated independently in each field and then combined. 
We report power spectra in terms of \dl, defined as 
\begin{equation}
  D_{\ell} = \frac{\ell (\ell + 1)}{2\pi} C_{\ell} \, .
\end{equation}
Throughout this work, we will often refer to the various single-frequency and cross-frequency spectra using
the shorthand ``\ninenine'' for the 95 GHz spectrum, ``\nineone'' for the 95 GHz -- 150 GHz
spectrum, etc.

%%%% Maps
%\subsection{Maps}
%\label{sec:maps}
%
%The cross-spectrum estimator takes as input a set of maps for each frequency and for every field. 
%For most fields, each map is a single observation. 
%The noise in each observation is statistically independent, given that the observations are taken at least an hour apart and the timestream data is high-pass filtered. 
%For the four fields observed with a lead-trail strategy, each map is a coadd of one lead and one trail observation. 
%The \one{} field was observed using larger elevation steps; we thus define an observation for this field as a coadd of two pairs of lead-trail observations with different elevation dithers, resulting in more uniform coverage.
%

%%%%
%\subsection{Window}
%\label{sec:window}
%
%Before taking Fourier transforms of the maps, we multiply each map by a window \textbf{W}, in order to smooth sharp edges at map borders, control overlap between neighboring fields, and remove power from bright point sources. 
%All observations of a given field are multiplied by the same mask for all observing frequencies.
%The window is a product of an apodization mask with a point source mask.
%The apodization mask is obtained by selecting the combined uniform coverage region for the three observing frequencies and smoothing the result with a Hann function.
%We mask point sources with a 150 GHz flux above 50 mJy. Each source is masked with a $5'$ radius disk with a Gaussian taper of width $5'$ outside the disk. 
%The point source masks remove $x\%$ of the total sky area.
%
%
%%%
%\subsection{Cross-Spectra}
%\label{sec:crossspectra}

We use a cross-spectrum estimator similar to that described in \cite{polenta05}; 
specifically, we cross-correlate maps of different observations of the same field.
These maps can be in the same observing band or different observing bands. 
We assume that the noise is uncorrelated between different observations, so the cross-spectra are free of noise bias.
We multiply the map of observation $A$ at frequency $\nu_1$ by a mask, zero-pad it to the same size for all fields, and calculate its Fourier transform $\tilde{m}^{A}(\nu_1)$, then we do the same
for observation $B$ at frequency $\nu_2$. The mask \textbf{W} rolls off the noisy edges of the field and
has apodized holes at the locations of bright point sources (sources with flux density above 50 mJy at 150~GHz).
We calculate the average cross spectrum between these two maps
within an $\ell$-bin $b$:
\begin{equation}
\label{eqn:ddef}
 \widehat{D}^{AB}_b(\nu_1,\nu_2)\equiv \left< \frac{\ell(\ell+1)}{2\pi}H_{\pmb{\ell}}Re[\tilde{m}^{A}_{\pmb{\ell}}(\nu_1)\tilde{m}^{B*}_{\pmb{\ell}}(\nu_2)] \right>_{\ell \in b}. 
\end{equation}
Here, $\pmb{\ell}$ is the two-dimensional $\ell$-space vector, and $H_{\pmb{\ell}}$ is a two-dimensional weight array that is described below. 
We average all cross-spectra $\widehat{D}^{AB}_b$ for $A \neq B$ to calculate a binned power spectrum $\widehat{D}_b$ for each field and frequency combination. We will often refer to binned power spectrum values $D_b$ as \textit{bandpowers}.
%All observations are assigned equal weight. 

Noise in the SPT-SZ survey maps is statistically anisotropic. Modes of a given angular frequency $\ell$ that oscillate perpendicular to the scan direction ($\ell_y \simeq \ell$; $\ell_x \simeq 0$) are noisier than modes that oscillate along the scan direction ($\ell_x \simeq \ell$; $\ell_y \simeq 0$).
To combine the power from different modes in each $\ell$ bin more optimally, we define a two-dimensional weight array $H_{\pmb{\ell}} (\nu_1, \nu_2)$ for each frequency combination:
\begin{equation}
\label{eqn:2dweight}
H_{\pmb{\ell}} (\nu_1, \nu_2) \propto \left [C_\ell^{\rm{th}} (\nu_1, \nu_2) + N_{\pmb{\ell}} (\nu_1, \nu_2) \right ]^{-2} \,,
\end{equation}
 where $C_\ell^{\rm{th}}$ is the theoretical power spectrum used in simulations, and $N_{\pmb{\ell}}$ is the two-dimensional calibrated, beam-deconvolved noise power. The noise power is calculated 
 %by differencing maps made from left-going and right-going scans.
for each field and frequency as the average two-dimensional cross-power spectrum of 100 
noise realizations for each frequency in the cross-spectrum pair. 
Noise realizations are created under the assumption of stationarity (statistically identical
noise in each individual observation) by splitting the individual observations into two halves,
multiplying one half by -1, and averaging all observations. Multiple semi-independent 
realizations are created by randomizing which individual observations go into each half.
We smooth the weight array and normalize it to the maximum value in each annulus. 
In addition to this smooth weighting, which is identical to the treatment in S13, we also mask a set
of modes in the spectra involving 95 GHz with $\pmb{\ell}$ values that correspond to frequencies of 1 Hz in the time-ordered
data. If these modes are not masked, we find anomalously low PTE values in one of the null tests 
described in Section~\ref{sec:systematicTests}.

The raw bandpowers $\widehat{D}_b$ are a biased estimate of the true sky power, $D_b$. 
The biased and unbiased estimates are related by
\begin{equation}
\widehat{D}_{b^\prime} \equiv K_{b^\prime b} D_b \, ,
\end{equation} 
 where the $K$ matrix accounts for the effects of the binning, windowing, TOD filtering, pixelization, and beams.
Following \cite{hivon02}, $K$ can be written as
\begin{equation}
\label{eqn:kdef}
K_{bb^\prime}=P_{b\ell}\left(M_{\ell\ell^\prime}[\textbf{W}]\,F_{\ell^\prime}B^{2}_{\ell^\prime}\right)Q_{\ell^\prime b^\prime},
\end{equation}
where $Q_{\ell^\prime b^\prime}$ is the binning operator and $P_{b\ell}$ its reciprocal,
$M_{\ell \ell^\prime}$ is the matrix describing the mode-coupling induced by the mask, and $B_\ell$ and $F_\ell$ are the 
beam and filter transfer functions described in Section~\ref{sec:obs}.

After performing the analysis described above, we are left with 19 sets of $6 N_\mathrm{bins}$ bandpowers, 
%and 19
%$6 N_\mathrm{bins}$-by-$6 N_\mathrm{bins}$
%covariance matrices, 
one per SPT-SZ survey field. The total $\ell$ range covered is 
$650 < \ell < 3000$ with $\Delta \ell = 50$, resulting in $N_\mathrm{bins} = 47$ bins.
The value of $\Delta \ell$ follows the choice in K11 and S13, chosen to minimize
bin-to-bin correlation in a typical-size field while preserving resolution of peaks in the CMB 
power spectrum.
We combine the bandpowers 
%and 
%covariance matrices 
from each field using area-based weights.
The combined bandpowers 
%and covariance matrix 
are given by
\begin{equation}
D_b (\nu_1, \nu_2) = \sum_{i}D_{b}^{i}(\nu_1, \nu_2) \ w^{i}
\end{equation}
%\begin{equation}
%\label{eqn:combCov}
%\textbf{C}_{bb^\prime} = \sum_{i}\textbf{C}_{bb^\prime}^{i}(w^{i})^2
%\end{equation}
where 
\begin{equation}
w^{i} = \frac{A^i}{\sum_{i}A^i},
\end{equation}
and $A^i$ is the area of field $i$. For notation convenience, we will subsequently
refer to the full $6 N_\mathrm{bins}$ unbiased bandpowers as $D_\alpha$, where
$\alpha$ runs over the $N_\mathrm{bins}$ bins as the fast index and the six
frequency combinations as the slow index.

Because we have used nearly identical analysis methods and data for the 
\oneone\ spectrum as used in S13, we can compare the S13 bandpowers 
to the \oneone\ bandpowers in this work as a check. Once we account
for the slightly different calibration in the
two analyses (the two calibrations differ by roughly 0.9\% in temperature),
the fractional difference in bandpowers is less than 1\% in most 
$\ell$ bins and less than 2\% everywhere. This residual difference is attributable
to small differences in cuts, apodization, and $\ell$-space weighting between the two 
analyses.\footnote{The \oneone\ bandpowers used here are derived from nearly identical
data and cut and weight settings as the bandpowers used in \cite{hou18}, 
but that work used the cross-spectrum
of a single pair of half-depth maps rather than the mean cross-spectrum of all
pairs of single-observation maps. The bandpowers used in \cite{hou18} were 
compared to the S13 bandpowers, and no evidence of inconsistency was found.}
 
%%% Bandpower covariance
\section{Bandpower Covariance Matrix}
\label{sec:cov}
The bandpower covariance matrix quantifies the uncertainty on the 
$6 N_\mathrm{bins}$ bandpowers, as well as the correlations among
uncertainties in different $\ell$ bins and frequency combinations.
It includes terms describing the contributions
from instrument and atmospheric noise, sample variance, and uncertainty in
our knowledge of the instrument beam and absolute calibration.
The bandpower covariance matrix $C_{\alpha \beta}$ is a set of 36  
$N_\mathrm{bins}$-by-$N_\mathrm{bins}$ blocks, with the six on-diagonal blocks 
corresponding to the bin-to-bin covariances within the  
six power spectra (\ninenine, \nineone, etc.),
and the off-diagonal blocks corresponding to the covariances between 
one frequency combination and another. The indices of $C_{\alpha \beta}$ run
over $\ell$ bins and frequency combinations, as defined in Section~\ref{sec:ps}.
We describe how we estimate each of the contributions
to the bandpower covariance matrix, and our method for numerically conditioning
the matrix, below.
\subsection{Estimating and Conditioning the Noise and Signal Terms}
\label{sec:condition}
In S13 and G15 and earlier SPT analyses, we split the estimation of the signal contribution
to the bandpower covariance matrix from the estimation of the noise and noise $\times$ signal
parts. 
The signal term, or sample variance, is calculated from the same (noise-free) simulations used
to estimate the filter transfer function, while the noise and noise $\times$ signal terms are 
estimated from the data, namely from the distribution of cross-spectrum bandpowers 
over all the map pairs used in Eqn.~\ref{eqn:ddef}.
%between 
%different observations at different frequencies. Noise variance is included only for the blocks that 
%share common maps (in other words, if the two cross-spectra have a frequency in common -- for instance, 95 x 150 with 150 x 220). 
As discussed in detail in Section~\ref{sec:sv}, we do not include
sample variance in estimating the bandpower covariance 
matrix for the fitting procedure described in that section, though we do include the 
noise $\times$ signal contribution. 
To accomplish this, we simply omit 
from the covariance the term estimated from noise-free simulations.

In the limit of full sky coverage, no gravitational lensing, no masking, and stationary Gaussian noise, the measurement of 
the power spectrum at every value of multipole $\ell$ is uncorrelated from the measurement at 
every other value of $\ell$, and the covariance matrix of our binned, unbiased estimate $D_b$
would be diagonal for a single frequency combination. Even in this limit, however, the noise between 
different combinations that share a frequency would be correlated, so we must include off-diagonal blocks in the bandpower covariance matrix. 
Additionally, the finite sky coverage, point-source mask, and non-Gaussian foregrounds 
induce correlations between bins in a single frequency combination, so we construct the full 
$6 N_\mathrm{bins}$-by-$6 N_\mathrm{bins}$ covariance matrix.

%Because the signal we are attempting to measure is a variance, there is a signal contribution to 
%the bandpower covariance matrix, often referred to as sample variance. We estimate the sample
%variance using the same simulations used to estimate $F_\ell$, the filter transfer function. We estimate
%the contribution to the bandpower covariance from noise using the variance over the many cross-spectra
%averaged to produce $D_b$ for each field and frequency combination (for more details on this 
%calculation, see \cite{lueker10}).
%
%%The bandpower covariance matrix quantifies the bin-to-bin covariance of the unbiased spectrum.
%%It is a sum of terms that account for the signal, noise, beam, and calibration covariance.
%The bandpower covariance matrix is a set of 36 square blocks, each block of size nbin by nbin, with the six on-diagonal blocks corresponding to the covariances of 95 x 95, 95 x 150, 95 x 220, ..., 220 x 220 GHz spectra. 
%
%The signal term, often called ``sample variance'', is calculated from the 400 simulations described in Section \ref{sec:transfer}. For each simulated sky, we calculate ...
%
%The noise term is estimated from the data, namely from the distribution of cross-spectrum bandpowers between different observations at different frequencies. Noise variance is included only for the blocks that share common maps (in other words, if the two cross-spectra have a frequency in common -- for instance, 95 x 150 with 150 x 220). 

The initial estimate of the bandpower covariance matrix is noisy for the off-diagonal elements, both 
within a frequency-combination block and in the off-diagonal blocks. To condition this noisy estimate, 
we compute the correlation matrices of the six on-diagonal blocks and note that their shape is determined 
primarily by the mode-coupling matrix and depends only on the distance from the diagonal. 
We condition the on-diagonal blocks by averaging the off-diagonal elements at a fixed separation 
from the diagonal and set elements that are a distance $\ell > 250$ from the diagonal to zero.
%since we do not expect any mode coupling beyond this scale.
This method of conditioning does miss some real correlation structure in the sample covariance. 
At very high $\ell$ in 
some frequency combinations, the covariance becomes dominated by the brightest point sources
and is hence strongly correlated across all $\ell$ bins. We account for this when we condition the 
sample covariance by fitting for and removing a fully correlated part of the covariance at high $\ell$
and replacing it after the conditioning described above. There are also potential off-diagonal terms
in the sample covariance from the effects of lensing; we expect these to be small for the temperature 
power spectrum and do not attempt to recover them in the conditioning. We note that the sample-variance
portion of the covariance matrix is omitted in the fitting procedure described in Section~\ref{sec:consist}, 
so the exact method of conditioning that part of the matrix does not affect the main result of this work.

For the off-diagonal blocks, the uncertainty on the computed covariance can be large compared to the true covariance. Therefore, we calculate the elements of the off-diagonal blocks by applying the average correlation matrix computed from the corresponding on-diagonal blocks to them.

We calculate and condition the sample and noise covariances on a by-field basis and then combine the results
using the same weighting used to combine bandpowers. The combined covariance matrix 
is given by
\begin{equation}
\label{eqn:combCov}
\textbf{C}_{\alpha \beta} = \sum_{i}\textbf{C}_{\alpha \beta}^{i} \ (w^{i})^2,
\end{equation}
where $w^i$ are the field weights defined in Section~\ref{sec:ps}.
%
%\subsection{Beam, Calibration, and Foreground-removal Contributions}
\subsection{Foreground-removal, Beam, and Calibration Contributions}
We add a contribution to the covariance matrix to account for the uncertainty in the foreground components which were subtracted to generate our CMB-only band powers.
See Section~\ref{sec:foregroundTreatment} for details on the foreground
removal process and the covariance estimate.

As in S13 and G15 and earlier SPT papers, we also add components to 
the covariance matrix to account for beam and calibration uncertainty. 
%Unlike earlier papers, 
%however, we create the beam and calibration covariance matrices using a theoretical prediction
%for the measured bandpowers rather than the measured bandpowers themselves. This eliminates
%a small bias in parameter fitting (described in \cite{aylor17}---we note that this modification is also
%used in the most recent publicly released likelihood for S13). 
We follow the treatment in \cite{aylor17} and create the beam and calibration covariance matrices using a 
theoretical prediction for the measured bandpowers rather than the measured bandpowers themselves.
For the theoretical prediction, we
use CMB only, because the additions to the covariance matrix from uncertainty in subtracted foreground 
power 
%(see Section~\ref{sec:foregroundTreatment} for details) 
include beam and calibration contributions.
We follow the beam uncertainty treatment described in detail in K11. 
%We separate the sources of beam 
%uncertainty into those expected to be highly correlated between bands and those expected to be
%uncorrelated, and we only include the uncorrelated sources in the covariance matrix for the 
%analysis in this work. 
We note that the calibration uncertainty at 150 GHz is uncorrelated from the calibration 
uncertainty in the other two bands because of the different method of estimation (comparison of
SPT maps to \planck\ maps in the SPT-SZ observing region for 150 GHz, comparison of SPT power 
spectra to full-sky \planck\ power spectra for 95 and 220 GHz, for details see Section~\ref{sec:obs}).
%, so we use the
%full calibration covariance in this work.

%Finally, we add a contribution to the covariance matrix of our estimate of the foreground-cleaned,
%CMB-only bandpowers that comes from the uncertainty in the foreground components subtracted
%from the raw bandpowers. See Section~\ref{sec:foregroundTreatment} for details on the foreground
%removal process and the covariance estimate.

%These are instead treated as
%free parameters in the fits described in Sections~\ref{sec:consist} and \ref{sec:cosmofit}. We do calculate what the
%beam and calibration contributions to the covariance matrices would have been (following the procedure
%described in Section~3.5 of S13) but only for the purpose of comparing relative importance of 
%different sources of sources of uncertainty
%%% Combining Fields
%\subsection{Combining Fields}
%\label{sec:combine}

%-------------------
% errors figure
%-------------------

\begin{figure*}
\begin{center}
    \includegraphics[width=0.49\textwidth]{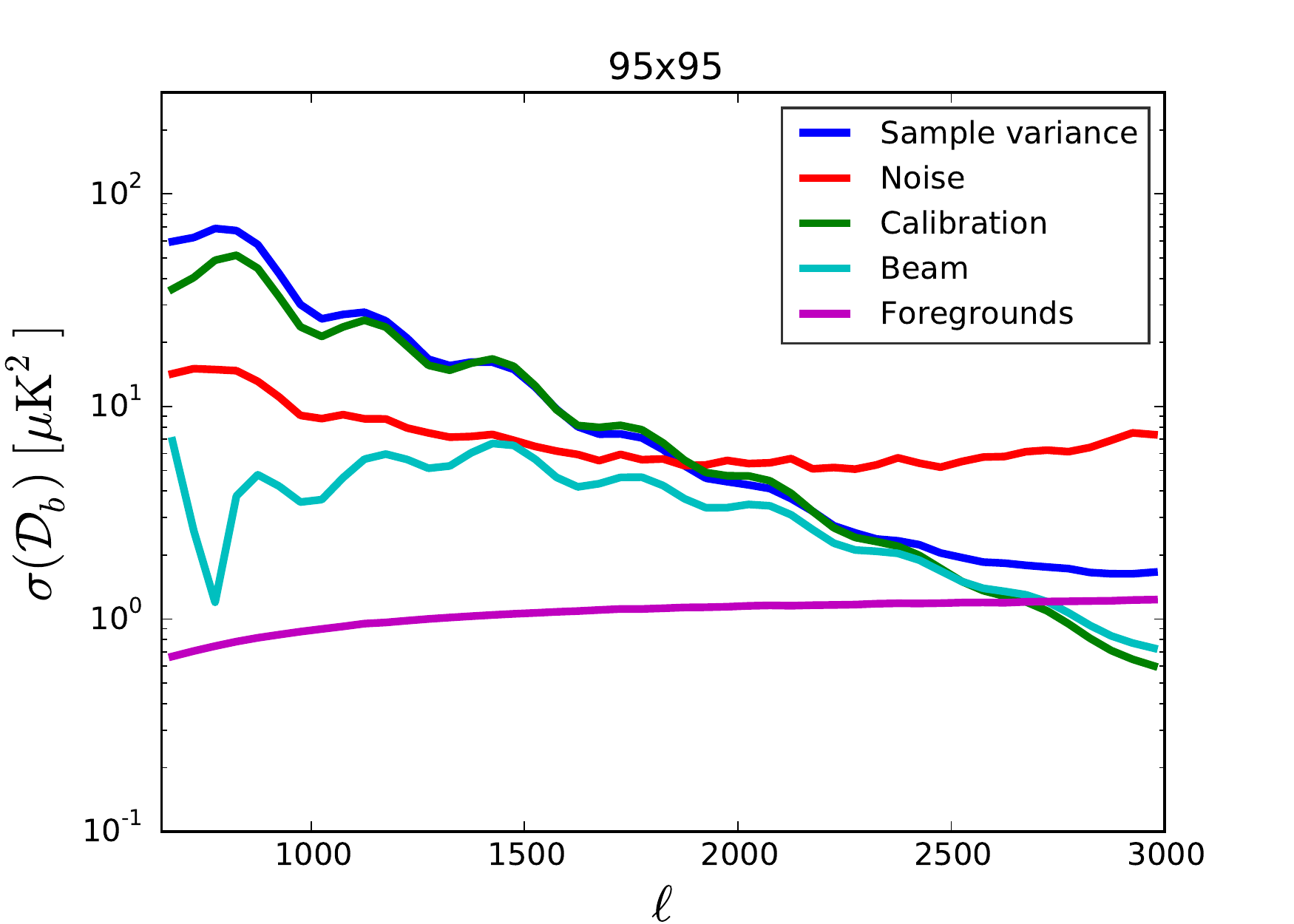}
    \includegraphics[width=0.49\textwidth]{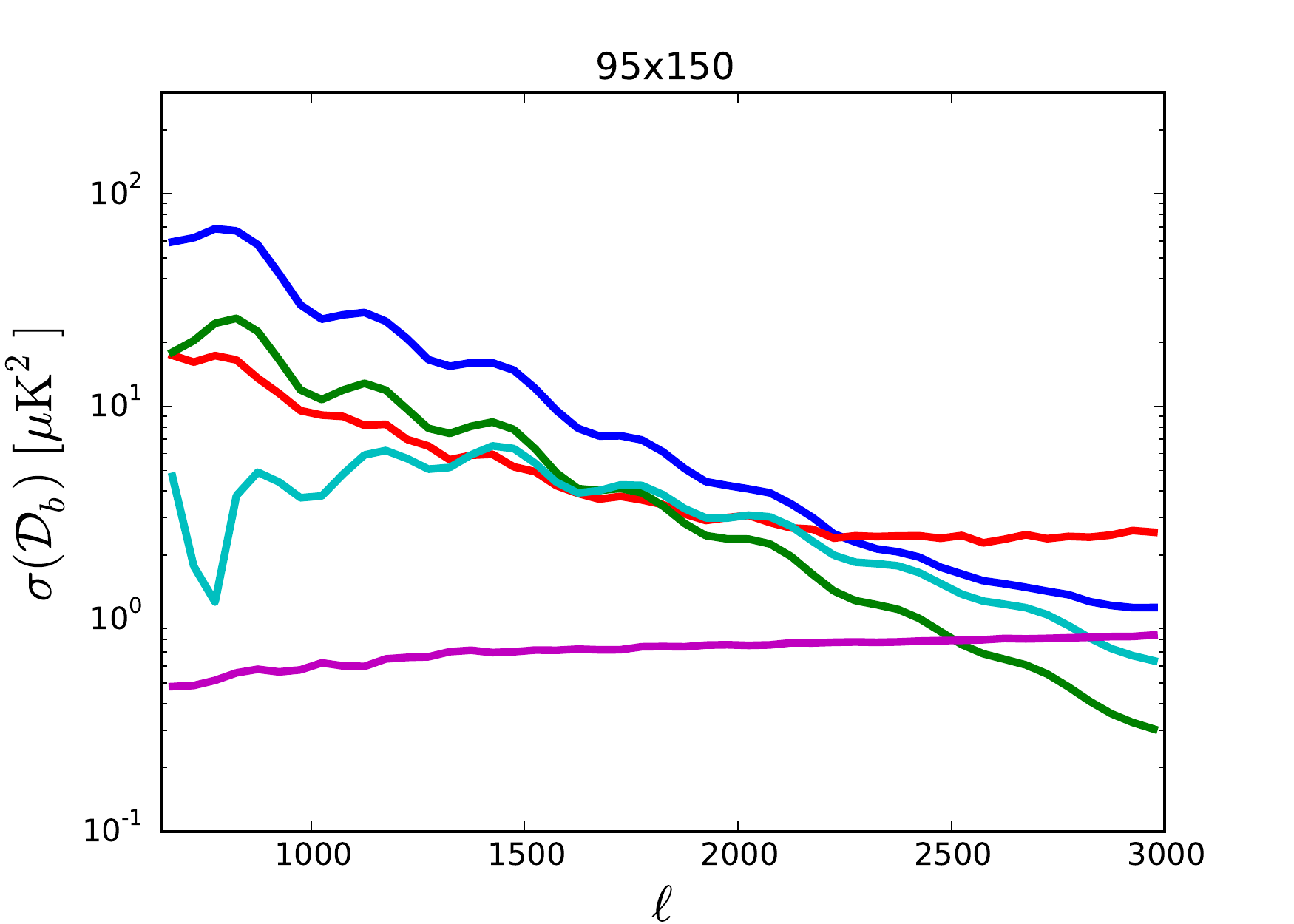}
    \includegraphics[width=0.49\textwidth]{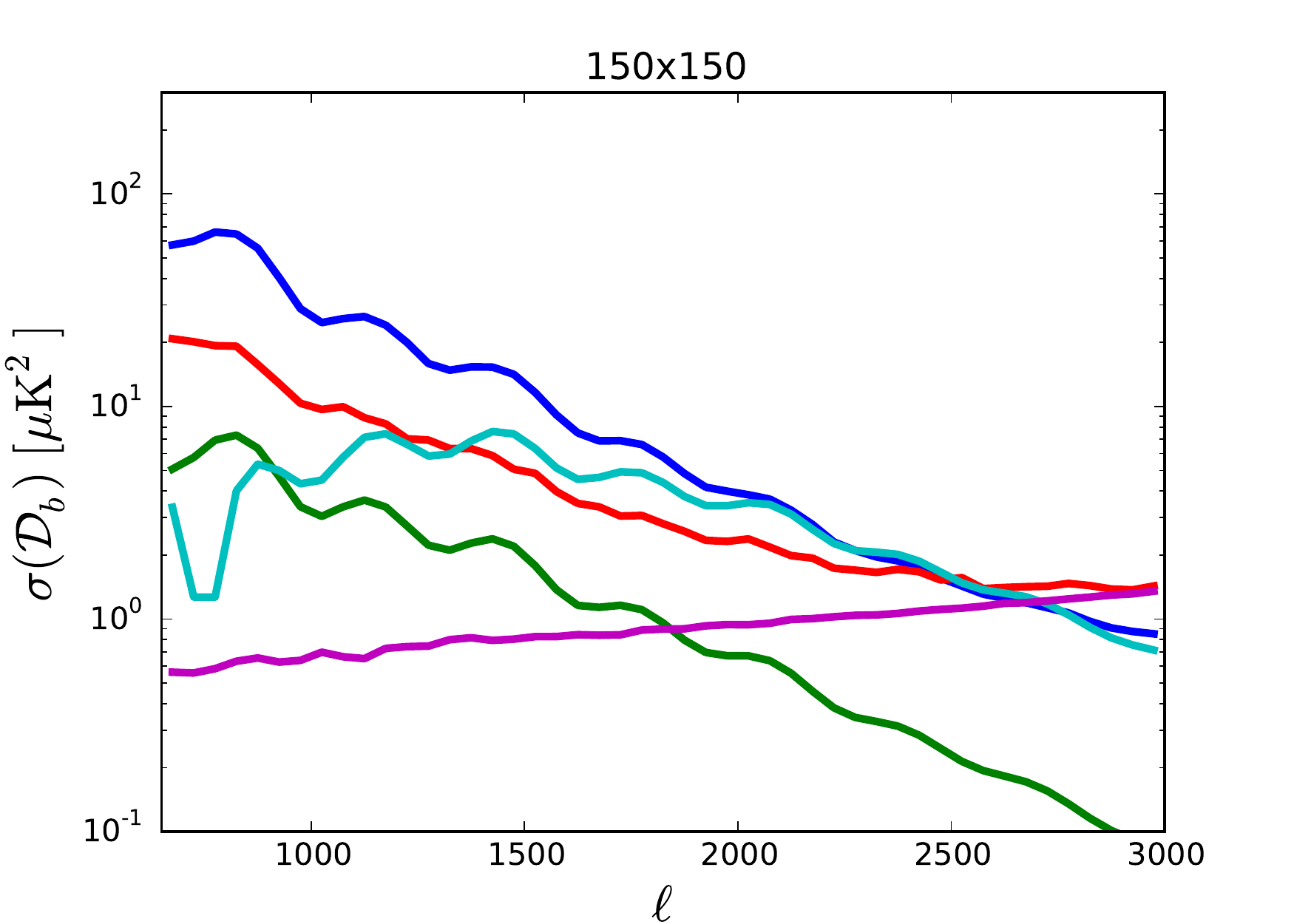}
    \includegraphics[width=0.49\textwidth]{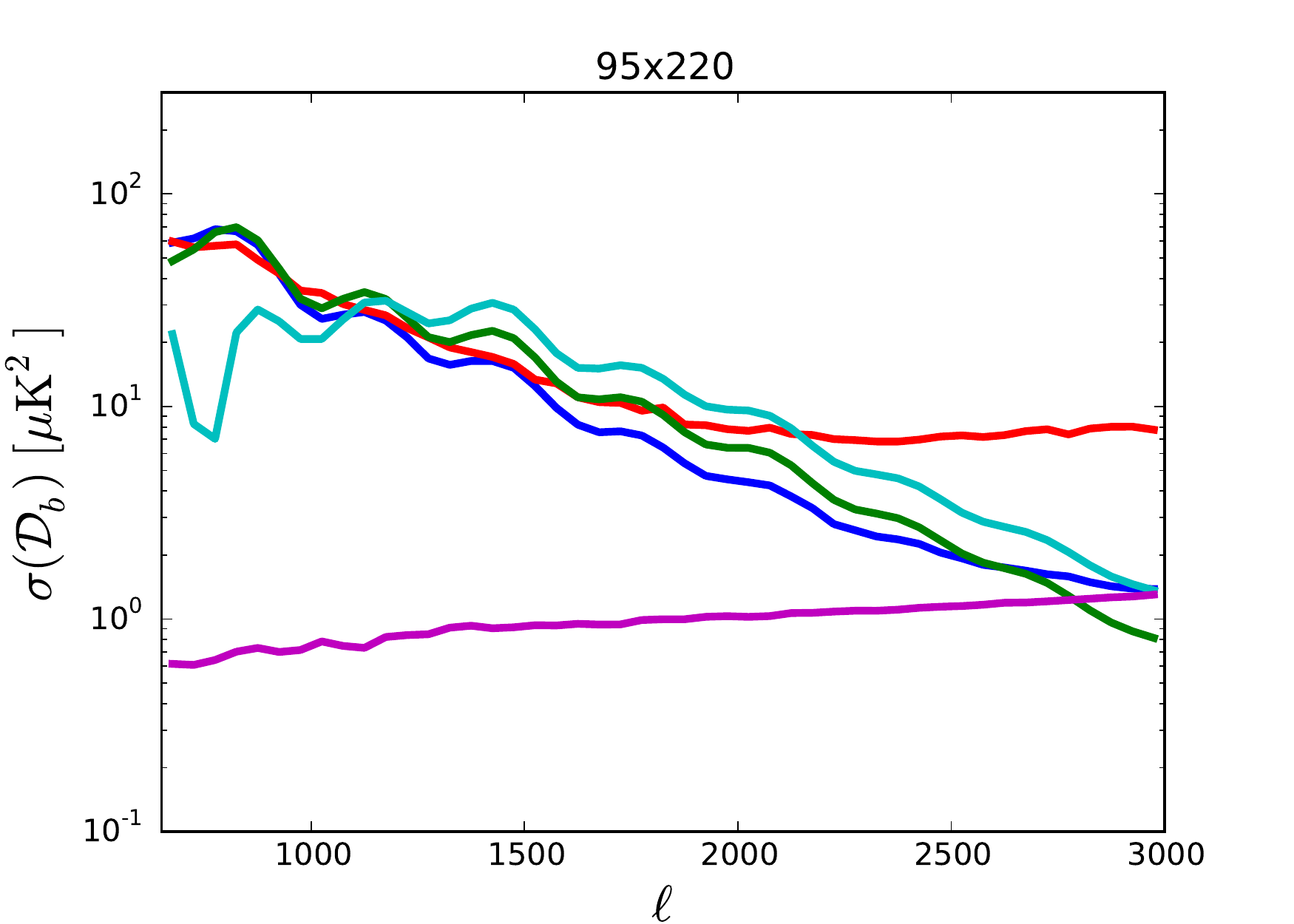}
    \includegraphics[width=0.49\textwidth]{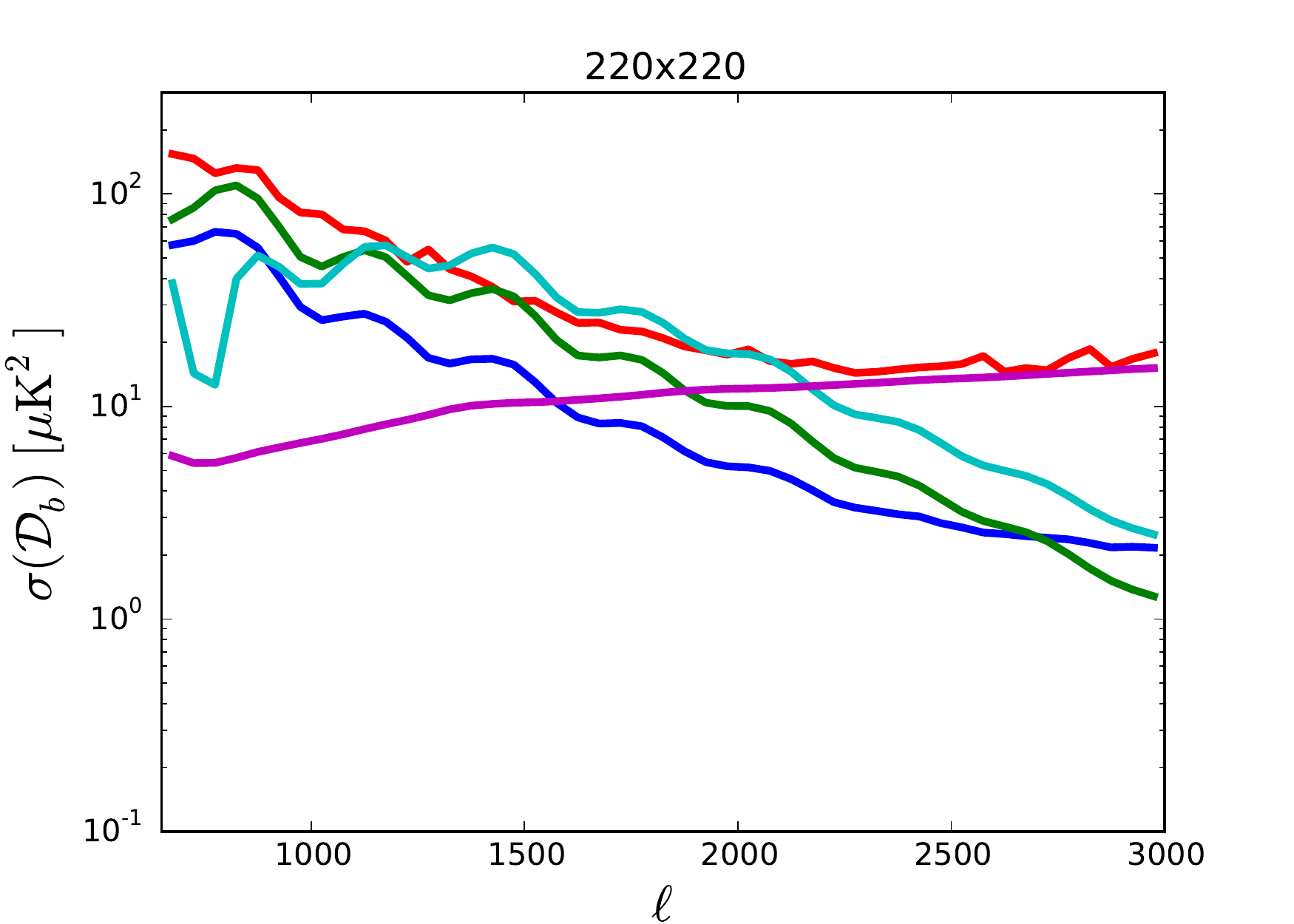}
    \includegraphics[width=0.49\textwidth]{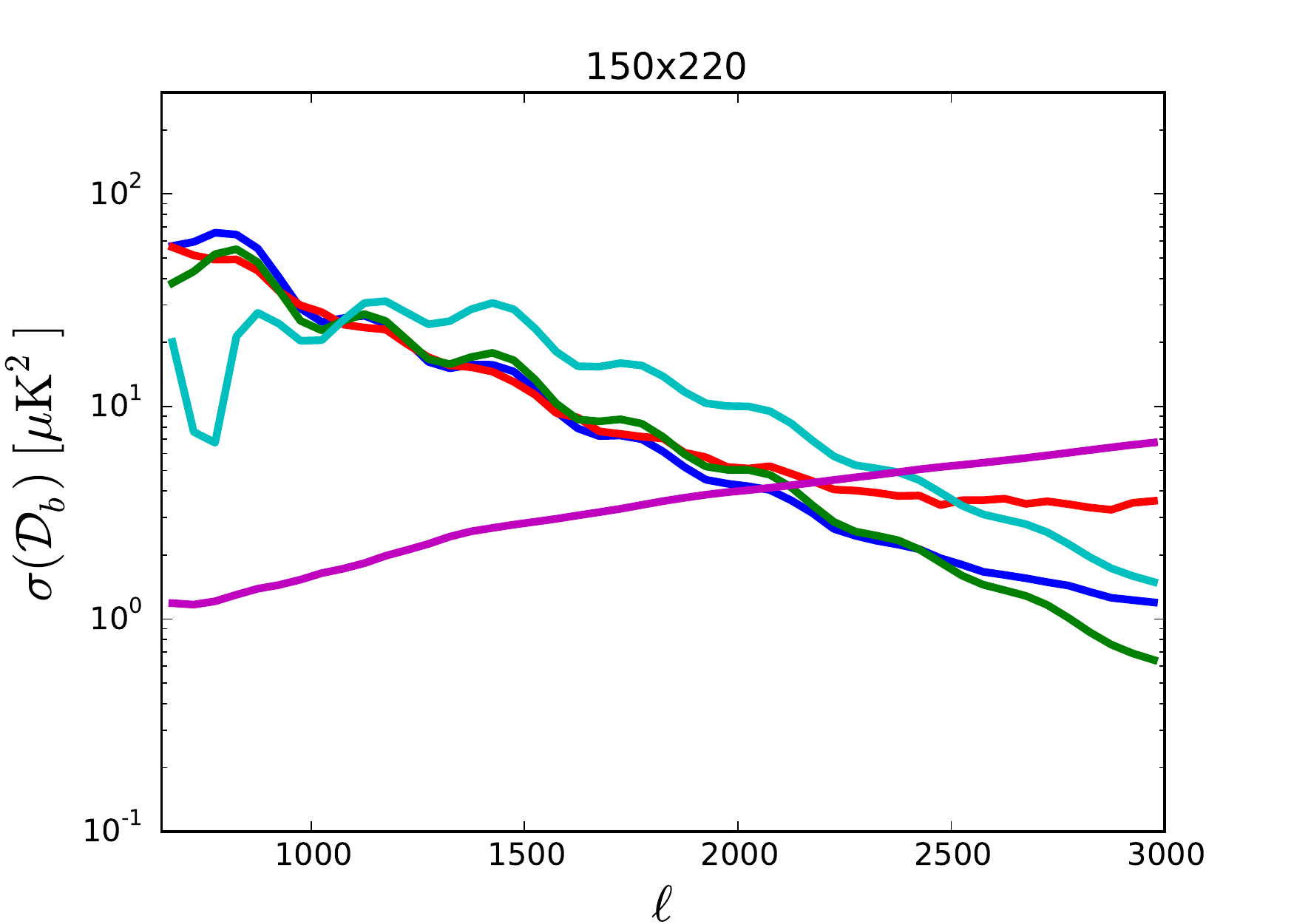}
\end{center}
\caption{
  Square root of the diagonal elements of the bandpower covariance matrix for
  each frequency combination, broken down into contributions from five different
  sources. We note that the sample variance contributions are
  not included in the fitting procedure described in Section~\ref{sec:consist}.
  The covariance matrices are binned at the same $\Delta \ell = 50$ resolution
  as the data, so the plotted diagonal values also have that resolution.
  }
\label{fig:errors}
\end{figure*}

\begin{figure*}
\begin{center}
    \includegraphics[width=0.49\textwidth]{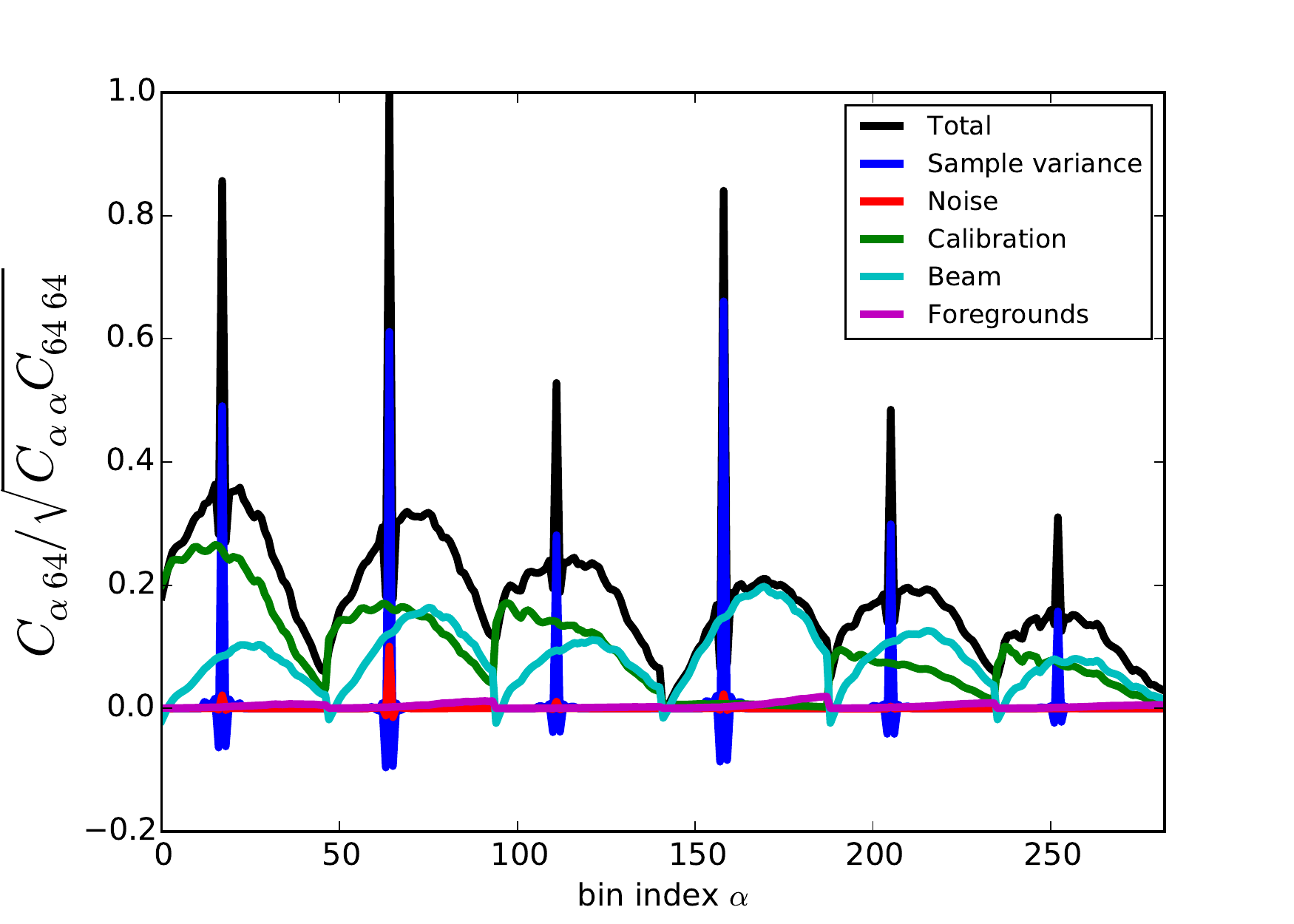}
    \includegraphics[width=0.49\textwidth]{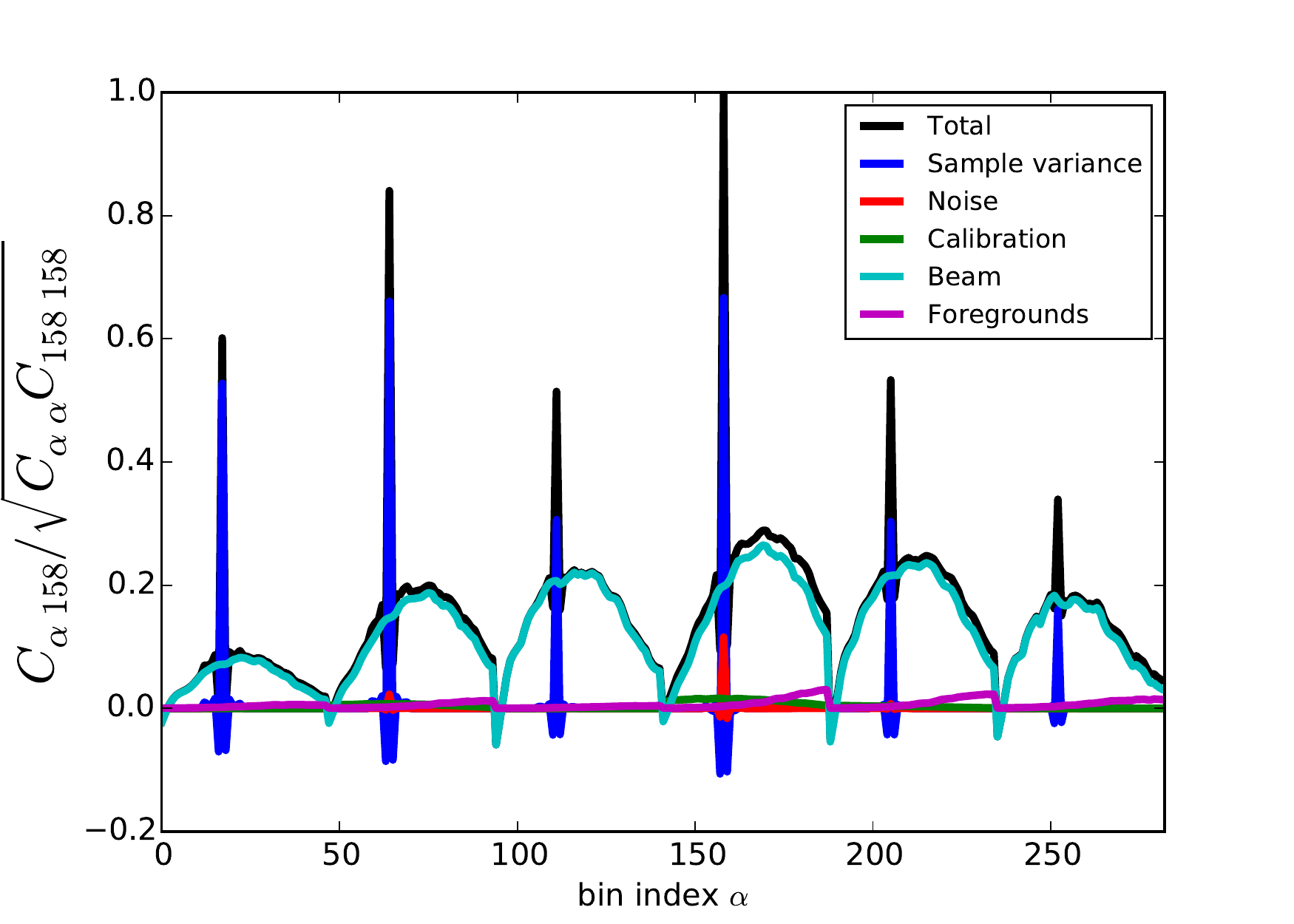}
\end{center}
\caption{
Slices through the covariance matrix along the rows corresponding to the $\ell=1525$
bin in \nineone\ (\textit{left panel}) and \oneone\ (\textit{right panel}). The $x$-axis corresponds
to bin index $\alpha$ (cf.~Section~\ref{sec:ps}), which runs over $\ell$-space bins from 
$\ell=650$ to $\ell=3000$ six times, once each for \ninenine, \nineone, etc. (so bin 
64 is $\ell=1525$ in \nineone, and bin 158 is $\ell=1525$ in \oneone).
All slices are normalized by the square
root of the product of the diagonal of the total covariance and the $C_{64 \; 64}$ or
$C_{158 \; 158}$ element. 
We note that the sample variance contributions are
not included in the fitting procedure described in Section~\ref{sec:consist}.
}
\label{fig:covslice}
\end{figure*}

\subsection{Relative Size of Covariance Matrix Contributions}
The (square root of the) diagonal parts of various contributions to the field-combined covariance 
matrices are shown in Figure \ref{fig:errors} for the six single-frequency and cross-frequency spectra.
The ``foreground'' curves are the sum of the contribution from uncertainties in the G15
foreground estimates and uncertainties on the estimates of power from point sources masked
in G15 and not in this work (see Section~\ref{sec:foregroundTreatment} for details).
While the foreground component only applies to the foreground-cleaned, CMB-only 
bandpowers, and we do not include sample variance in the consistency test described
in Section~\ref{sec:consist}, we include both the sample variance and the foreground 
components in the figure for informational purposes. 

It is also worth mentioning some of the 
correlation structure of the various parts of the covariance matrix. In the regions of $\ell$
space in which the CMB is the largest source of signal ($\ell \lesssim 2500$ or 3000, depending 
on frequency combination---see Figure~\ref{fig:dlraw}), the sample variance is highly correlated
between frequency combinations but mostly uncorrelated between $\ell$ bins. At very high $\ell$
in some frequency combinations (particularly \ninenine), the sample variance becomes dominated
by the brightest point sources and is thus highly correlated between $\ell$ bins and, to a degree,
among frequency combinations. The noise variance is mostly uncorrelated between bins and 
frequency combinations.
The calibration covariance contributions are 100\% correlated between bins and 
partially correlated between frequency combinations (particularly among \ninenine, 
\ninetwo, and \twotwo\ because of the method used to calibrate the 95 GHz and 220 GHz 
data, described in Section~\ref{sec:obs}). The
beam contributions are strongly correlated between bins, strongly correlated between
frequency combinations that share a band,
and partially correlated between
frequency combinations that do not share a band.

As representative examples, slices through the covariance matrix along the rows
corresponding to the $\ell$-space bin centered at $\ell=1525$ in the 
\nineone\ and \oneone\ spectra ($C_{\alpha 64}$ and $C_{\alpha 158}$, 
respectively, in the numbering scheme introduced above) are plotted in Figure~\ref{fig:covslice}.
We choose these two spectra because they have the lowest total covariance
and thus the highest weight in the fit described in Section~\ref{sec:consistall}.
We show slices through the total covariance and through the sample variance, noise, calibration, beam,
and foreground components individually. All slices are normalized by the square
root of the product of the diagonal of the total covariance and the $C_{64 \; 64}$ or
$C_{158 \; 158}$ element. 

\section{Null Tests}
\label{sec:systematicTests}

While the purpose of this analysis is to test for unmodeled systematics in 
the final reported SPT-SZ bandpowers by comparing the different single-frequency and
cross-frequency spectra, we can also check the individual 
spectra for contamination by splitting the data along some axis that is expected
to maximize the sensitivity to a known source of contamination, calculating the
power spectrum of the differenced data, and comparing to the expected value
of such a difference from simulations.
%We test for instrumental systematics in the same way as S13, G15, and previous SPT power
%spectrum analyses, by performing a series of null tests on data splits. 
We follow S13, G15, and previous SPT analyses in performing these null tests.
Briefly, in each null test, we divide a set of maps into two halves 
and subtract one set from the other, removing most of the signal (some residual
signal is expected due to slight differences in weights, filtering, and calibration
in individual observations). We then repeat the cross-spectrum analysis described
in Section~\ref{sec:ps} but with the subtracted ``null maps,''
%of the set of differenced maps
and we compare the result to the expectation value of residual signal, using the 
%bandpower covariance matrix described in the previous section 
variance over many cross-spectra to test whether
any difference from the expectation value is consistent with noise.

The data splits are chosen
to maximize the sensitivity to certain families of systematic uncertainties typical of ground-based
CMB data. In this work, we use the ``Time,'' ``Scan Direction,'' ``Azimuthal Range,'' and ``Sun''
tests defined in S13. We perform them on the 95, 150, and 220 GHz data individually, 
for a total of 12 tests. The PTEs from these tests are shown in Table~\ref{tab:null}.
They range from 2\% to 78\%, three of the 
PTE values are below 10\%, and ten of the PTE values are below 50\%. 
%If the twelve tests
%were fully independent, 
In the absence of significant contamination, 
we would expect a minimum PTE value among the 12 tests of 2\% or less roughly 20\% of the
time, three values below 10\% roughly 10\% of the time, and a distribution this asymmetric
roughly 2\% of the time. This takes into account the correlation among the null test values
expected from the fact that they are not perfectly orthogonal splits of the data. (The largest
expected correlation is 15-20\% between the Time and Sun splits.)
%In reality, the significance of any of these a posteriori statistics is 
%lower than this (i.e., we would expect the observed distribution more often), because the 
%twelve tests are mildly positively correlated.
%, and their distribution
%is marginally consistent with uniform (K-S test probability of 1.1\%). 
We make the subjective decision that 1-2$\sigma$ fluctuations in 
statistics defined a posteriori are not 
%In the absence of 
strong evidence of contamination, and 
%in the data from the individual frequency bands, 
we proceed to subtract an estimate of foregrounds from
the data and perform the consistency test.

\begin{table}[tbp]
\centering
\begin{tabular}{| l | c | c | c | c |}
\hline
Band  & Left-right & 1st-half-second-half & sun & azimuth \\
\hline
95~GHz & 0.08	 & 0.02 & 0.19 & 0.07 \\
150~GHz & 0.44 & 0.51 & 0.78 & 0.1 \\
220~GHz & 0.49 & 0.13 & 0.22 & 0.14 \\
\hline
\end{tabular}
\caption{\label{tab:null} 
Probability to exceed (PTE) the \chisq\ value obtained in each null test
in each band. The individual tests are not independent; tests of 
the overall distribution of PTEs are discussed in the text.}
\end{table}

%%%%%%%%%%%%%%%%%%%%%%%%%%%%%%%%%%%%
% FOREGROUND TREATMENT
%%%%%%%%%%%%%%%%%%%%%%%%%%%%%%%%%%%%
\section{Foreground treatment}
\label{sec:foregroundTreatment}

Our bandpowers include contributions from primary CMB temperature anisotropies and several foreground components. To obtain primary CMB-only bandpowers, we subtract estimates for foreground
bandpowers from our data. 
We also add a contribution to the bandpower covariance matrix (Section~\ref{sec:cov}) to account for the uncertainty in the subtracted foregrounds.

%\subsection{High-$\ell$ foregrounds}
%\label{section:fg_highell}

We adopt the baseline model from G15 to estimate foreground contributions to these bandpowers.
 The baseline model includes contributions from the 
thermal and kinematic Sunyaev-Zel'dovich effects (tSZ and kSZ), 
spatially clustered and unclustered contributions from the cosmic infrared background (CIB), 
and a contribution from spatially unclustered radio galaxies (i.e., galaxies emitting synchrotron radiation from an active galactic nucleus). 
In the baseline model, eight parameters are derived for those foregrounds: the amplitudes of tSZ, kSZ power, unclustered CIB power, and clustered CIB power; two parameters describing the frequency dependence of the CIB terms; the tSZ-CIB correlation; and the spectral index of radio galaxies.
The amplitude of radio galaxy power is fixed to a prior value based on the 
model in \cite{dezotti05}.
The analysis in G15 used the same raw data as this analysis; 
we refit the foreground model cutting G15 data at $\ell<3000$, so that the foreground constraints are independent of the data in this work. 

%\subsection{Masked point source power}
%\label{section:fg_ps}
The analysis in G15 used a point source mask with a threshold of 5$\sigma$ at 150 GHz, corresponding to a flux density of roughly 6.4 mJy at our survey depth. 
This analysis uses a masking threshold of 50 mJy. 
Consequently, the best-fit foreground model from our reanalysis of G15 data does not include 
the contribution from point sources between these two thresholds (roughly 3000 sources
total, or roughly one per square degree and 150 per individual field), and we need to estimate 
and remove the power from these sources separately. 
%We use two different approaches to this estimation and carry the 
%analysis through to the final result for both approaches.

%In the first approach, 
To estimate the power from sources between the two 
masking thresholds, we make use of the fact that we have full posterior deboosted flux density
distributions for every source masked in G15 but not masked in this work. These distributions are
calculated as part of the source count analysis in \cite{vieira10,mocanu13}
and W. Everett et al. (in preparation). 
We take 50,000 mock source catalogs 
drawn from these posterior flux distributions and calculate the contribution to the cross-spectrum 
between observing frequencies $\nu_1$ and $\nu_2$ from each of those mocks as:
\begin{equation}
\label{eqn:ptsrc}
\left(C_{\ell}^{\rm PS}\right)_{i, m}^{\nu_1\times\nu_2} = \frac{1}{A_i}\left(\frac{dB_{\nu_1}}{dT}\frac{dB_{\nu_2}}{dT}\right)^{-1}_{\rm T_{\rm CMB}}\sum_j S_j^{\nu_1}S_j^{\nu_2},
\end{equation}
where $i$ is the field index (19 fields total), $A$ is the area of the field in steradians,
$m$ is the mock catalog index, and $S_j^{\nu}$ is the flux of source $j$ (roughly 150 sources per 
individual field) from the list of selected sources in mock catalog $m$ at observing frequency $\nu$.
We then estimate the power from these sources $\left(C_l^{PS}\right)_i^{\nu_1\times\nu_2}$ in each field as the average of the power from all mocks and derive error bars on this quantity from the standard deviation of power among the mocks.
We subtract the resulting point source power in every field.

%The first 
This approach constitutes our best possible estimate of the power from the sources we intend
to mask, but it does not account for power from sources unintentionally masked. 
We do not expect a significant amount of power from unintentionally masked sources, because
sources at millimeter wavelengths between flux densities of 6 and 50 mJy are nearly all 
radio galaxies, and the clustering
of these sources at these flux density levels has been measured to be very small (e.g., G15).
%Because the bright
%point sources in the SPT catalog are generally massive galaxies, we expect an overdensity of less 
%(but not negligibly) bright sources around them. 
Nevertheless, as a cross-check we take a 
second approach to estimating the difference in
source power between G15 and this work. In this approach, we extend the multipole range in this analysis to 
$\ell_\mathrm{max}=4000$, we directly measure the difference in power between G15 and this work
in the range $3000 < \ell < 4000$, we fit the result to a constant $C_\ell$ spectrum, and we subtract
this value from the combined bandpowers in each single-frequency and cross-frequency combination. As in the
first approach, we also add a contribution to the covariance matrix from the uncertainty in this fit.
The final results using this approach are statistically indistinguishable from those obtained using our 
primary method.

\section{Inter-spectrum consistency test}
\label{sec:consistall}
\subsection{Formalism}
\label{sec:consist}
Our primary goal in this paper is to test the consistency among the 
six sets of foreground-cleaned bandpowers. 
We wish to perform this test independently of any assumptions about a cosmological model, so we choose to find the 
one number for each of the 47 $\Delta \ell=50$ bins which best fits the data, and we then evaluate the goodness 
of fit for the six bandpower sets to this computed best-fit set.
We apply a generalized least-squares fit to the six sets of bandpowers, obtaining a single set of 
best-fit bandpowers.  

We create a design matrix $A_{\alpha b}$ consisting of $N_\mathrm{bins} = 47$ vectors, each 
$6 N_\mathrm{bins}$ elements long. Each vector in the design matrix is equal to 1 in the six elements
corresponding to the measurement of the power in a single $\ell$-space bin in the six 
spectra and 0 otherwise. Our best estimate of the true CMB variance on our patch of the
sky in one bin is then
\begin{equation}
\label{eqn:dlbest}
\bar{D}_b = \left (A^T_{b \alpha} W_{\alpha \beta} A_{\beta b^\prime} \right )^{-1} A^T_{b^\prime \gamma} W_{\gamma \delta} D_\delta, 
\end{equation}
where (as defined in Section~\ref{sec:ps})
Greek indices run over the full $6 N_\mathrm{bins}$ bandpowers while 
Roman indices run over the $N$ best-fit bandpowers, 
$W_{\alpha \beta} = C^{-1}_{\alpha \beta}$ is the inverse bandpower covariance matrix (see
Section~\ref{sec:cov} for details), 
and summation over repeated indices is assumed.
Assuming Gaussian-distributed likelihoods and a covariance matrix that does not depend on 
the model, the probability of obtaining our data given
the model (which is simply that all six power spectra consist of a 100\% common signal 
and noise described by our covariance matrix) is $\mathcal{L} = e^{-\chisq/2}$, where
\begin{equation}
\label{eqn:chisq}
\chisq = (D_\alpha - M_\alpha)^T  \ W_{\alpha \beta} \ (D_\beta - M_\beta),
\end{equation}
and $M_\alpha = A_{\alpha b} \bar{D_b}$. The value of \chisq\ and the associated 
PTE are the primary results of this paper.

\subsection{Treatment of Sample Variance}
\label{sec:sv}
When comparing measured bandpowers 
to a cosmological model, we are interested in the true, underlying variance of the 
CMB anisotropy. This true variance is impossible to measure perfectly with a finite
number of samples, so the bandpower covariance matrix used in fitting to cosmology is the sum
of contributions from measurement uncertainty and the variance in the signal itself
(and the cross-terms). When comparing two different measurements of variance, however, 
the contribution of signal or sample variance to the covariance matrix of the comparison
depends on the fraction of sky modes the two measurements
have in common. In the limit that the same modes on the sky are used 
with the same weighting to estimate bandpowers, and the CMB is the only signal on the sky
(or the two measurements are at the same observing frequency),
the sample variance disappears entirely. Put another way, the fitting procedure described
in the previous section is formally insensitive to any component of the covariance 
matrix that is perfectly correlated between the six single-frequency and cross-frequency spectra.

We expect some residual uncorrelated sample variance among the six power spectra
measured in this work for two reasons: 1) at high $\ell$, the sky signal begins to be
dominated by foregrounds, not CMB, and the intensities of the various foreground sources
will be different in different bands; 2) the weighting of modes in an $\ell$-space bin 
(Eqn.~\ref{eqn:2dweight}) is slightly different for each spectrum.
The relative contribution of this residual uncorrelated sample variance to the final covariance
matrix should be small, though, for several reasons. One is that the CMB dominates
all contributions to foreground power in G15 even at the highest 
multipoles measured in this work ($\ell=3000$) for all spectra except
\twotwo, and that spectrum is noise- and beam-error-dominated at all multipoles
(see Figure~\ref{fig:errors}). The extra unclustered point-source power in the data used in this work
compared to G15 (see Section~\ref{sec:foregroundTreatment} for details) does overtake
the CMB power at $\ell < 3000$ in \ninenine\ (see Figure~\ref{fig:dlraw}), 
but the procedure used to 
remove this power (which derives from knowledge of the actual flux densities of individual
sources contributing to this power) also removes the sample variance. Furthermore, the 
sample variance on this signal is highly correlated between bins, because it is dominated by
the brightness of the brightest few sources in the map, so the process of removing this power by 
measuring it at $\ell > 3000$ will also remove the bulk of the sample variance. Finally, the 
difference in mode weighting is empirically found to be very small across the six power spectra. 

All of these effects can be properly taken into account in constructing the sample covariance
matrix, but we have empirically found that the large correlated component causes numerical
instability in the inversion of the full matrix, even after the conditioning described in 
Section~\ref{sec:condition}. For this reason, and based on the arguments that the residual 
uncorrelated sample variance should be small, we choose to ignore sample variance entirely 
in the fit in Section~\ref{sec:consist}. This will result in a small underestimate of the true variance
and make the consistency test marginally more difficult to pass.

%%%%%%%%%%%%%%%%%%%%%%%%%%%%%%%%%%%%
% BANDPOWERS AND CONSISTENCY
%%%%%%%%%%%%%%%%%%%%%%%%%%%%%%%%%%%%
\section{Results}
\label{sec:results}

%-------------------
% bandpowers figure 1
%-------------------

\begin{figure*}
\begin{center}
    \includegraphics[width=0.49\textwidth]{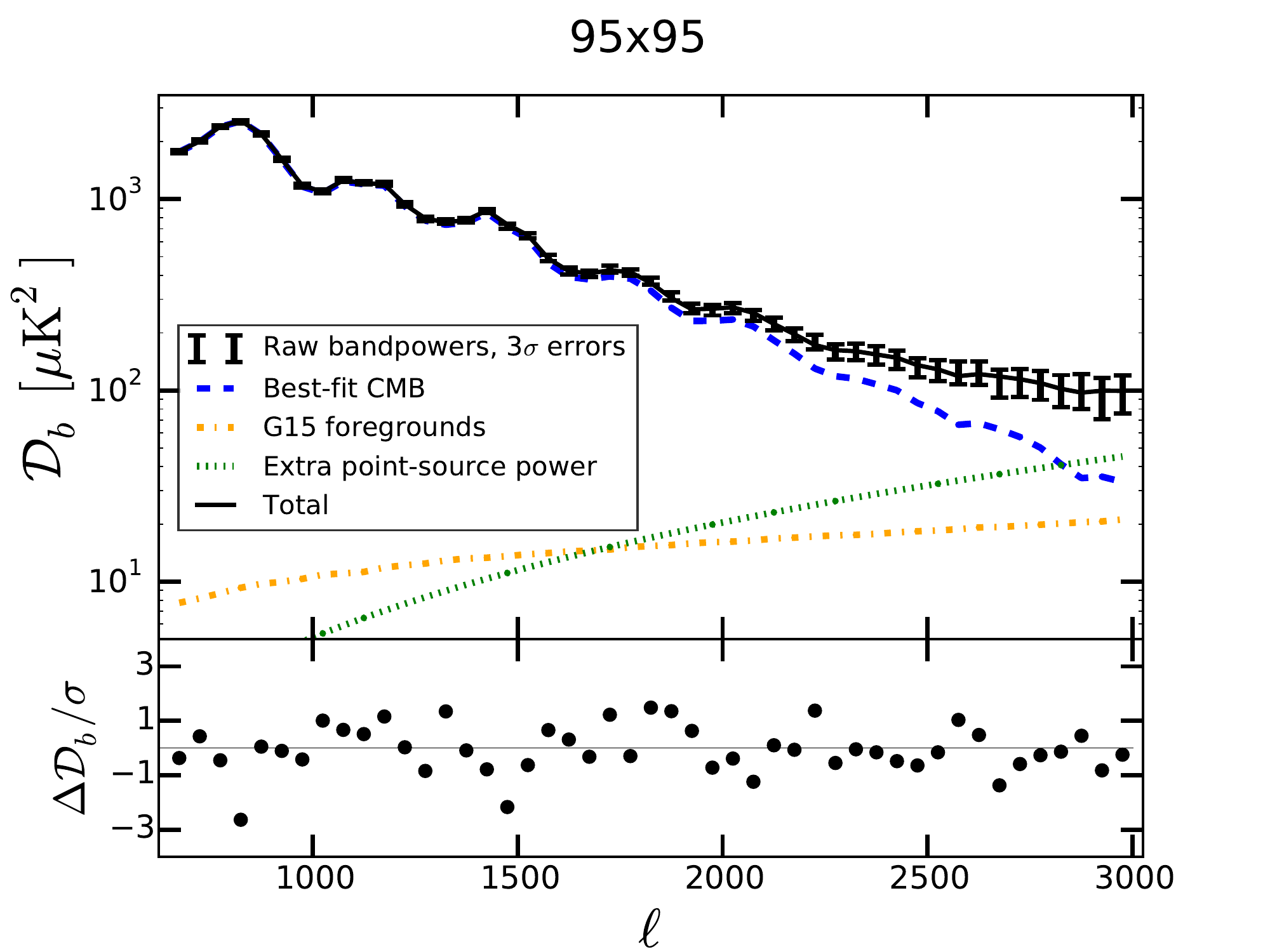}
    \includegraphics[width=0.49\textwidth]{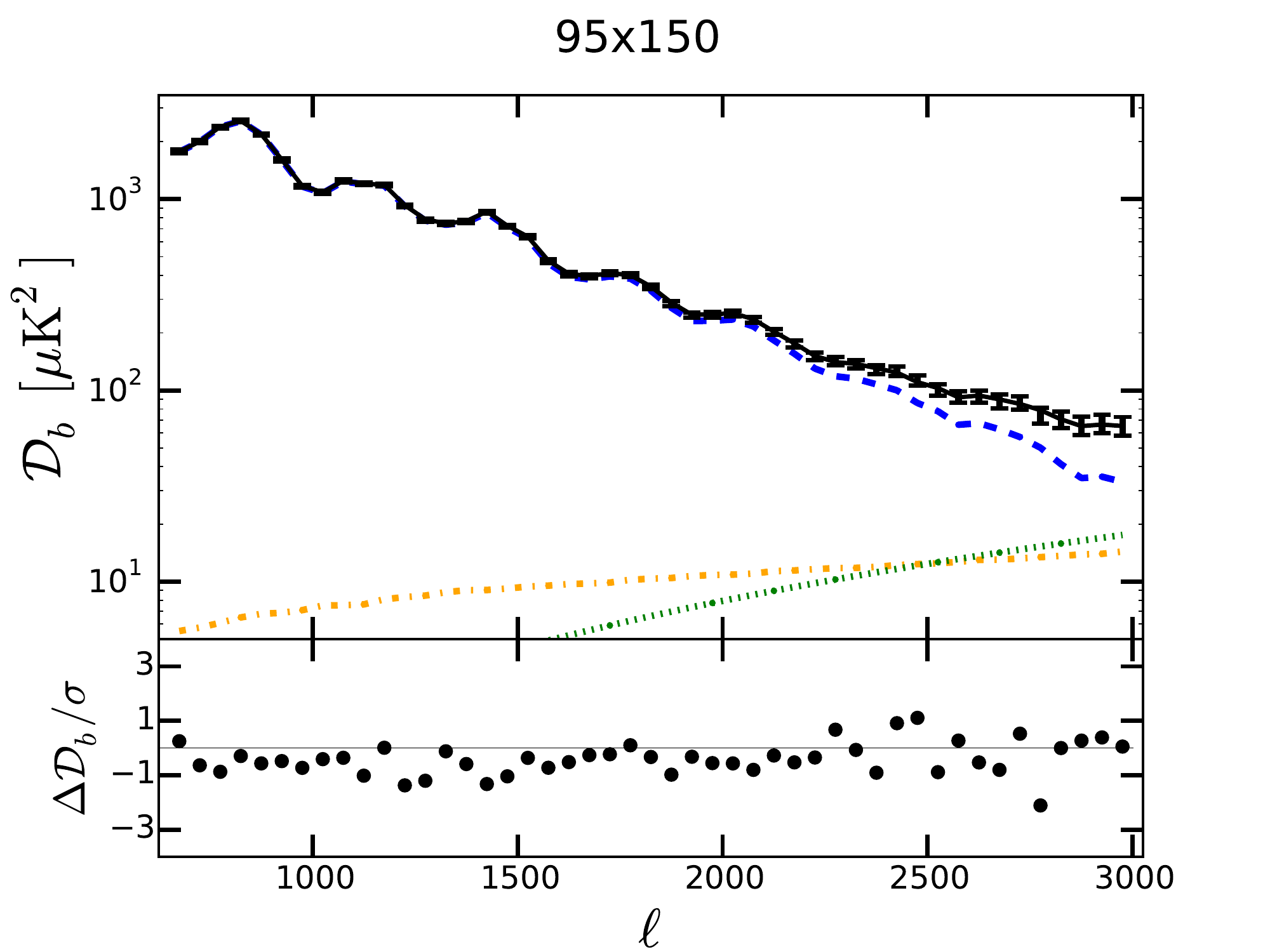}
    \includegraphics[width=0.49\textwidth]{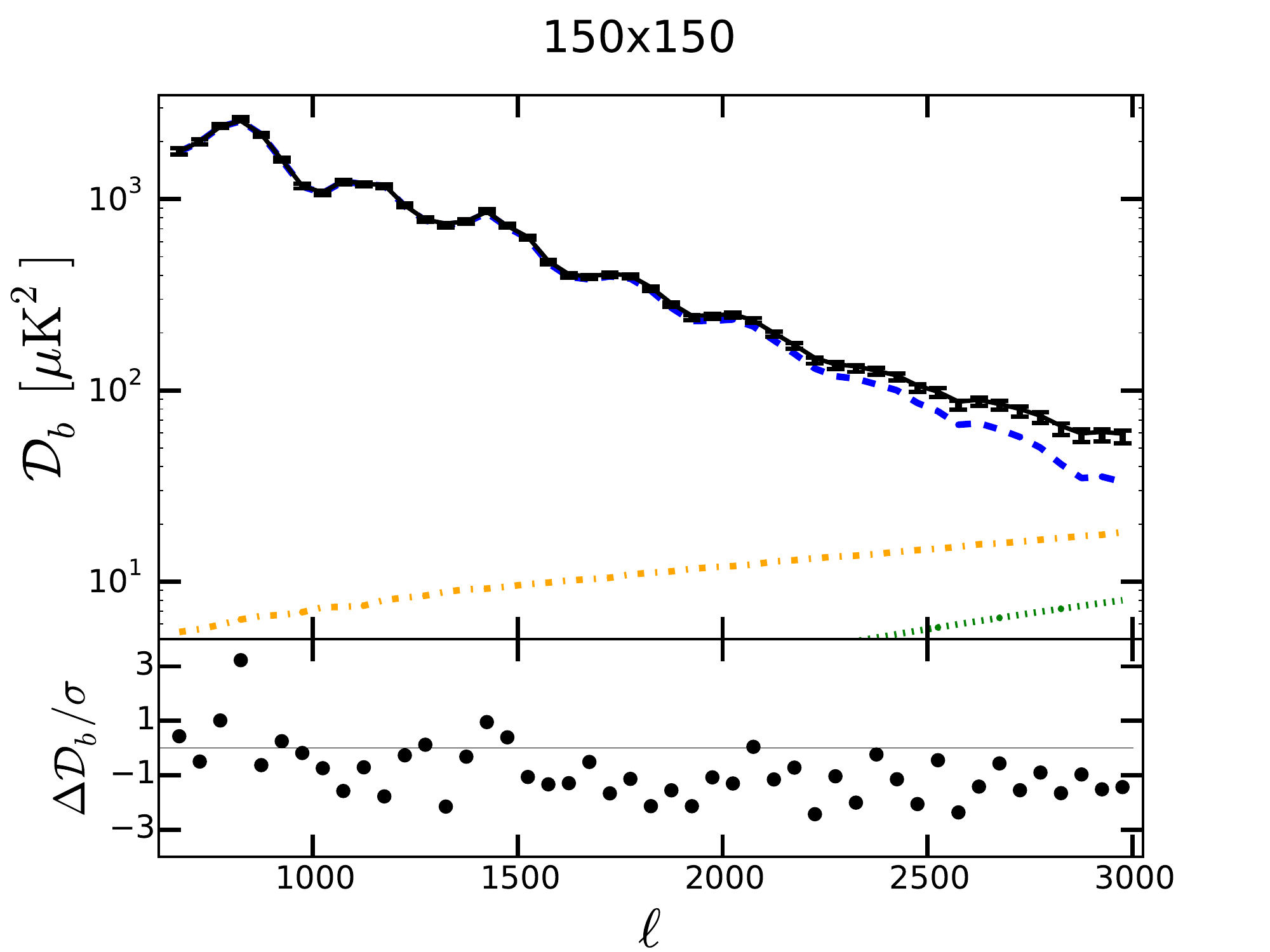}
    \includegraphics[width=0.49\textwidth]{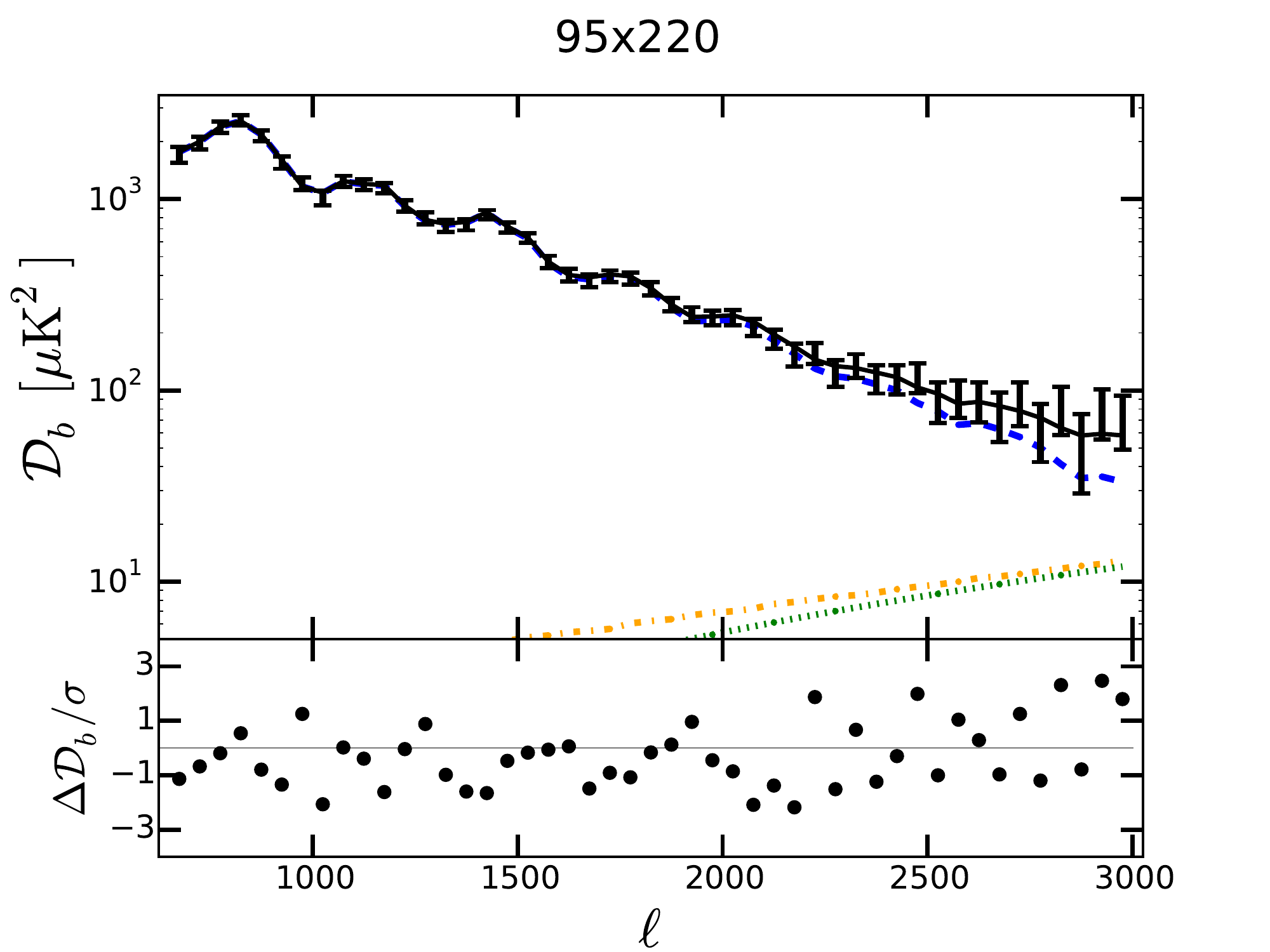}
    \includegraphics[width=0.49\textwidth]{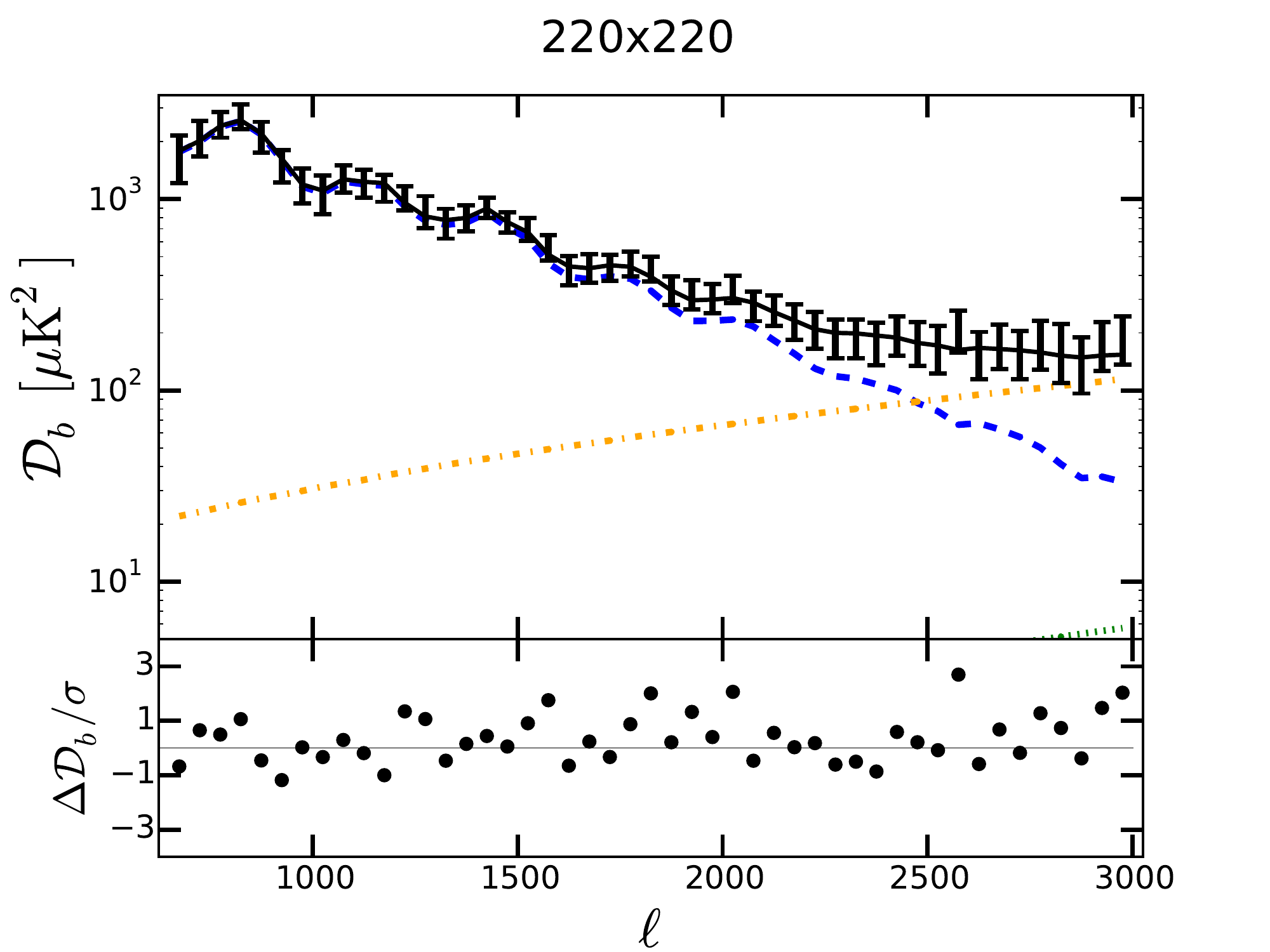}
    \includegraphics[width=0.49\textwidth]{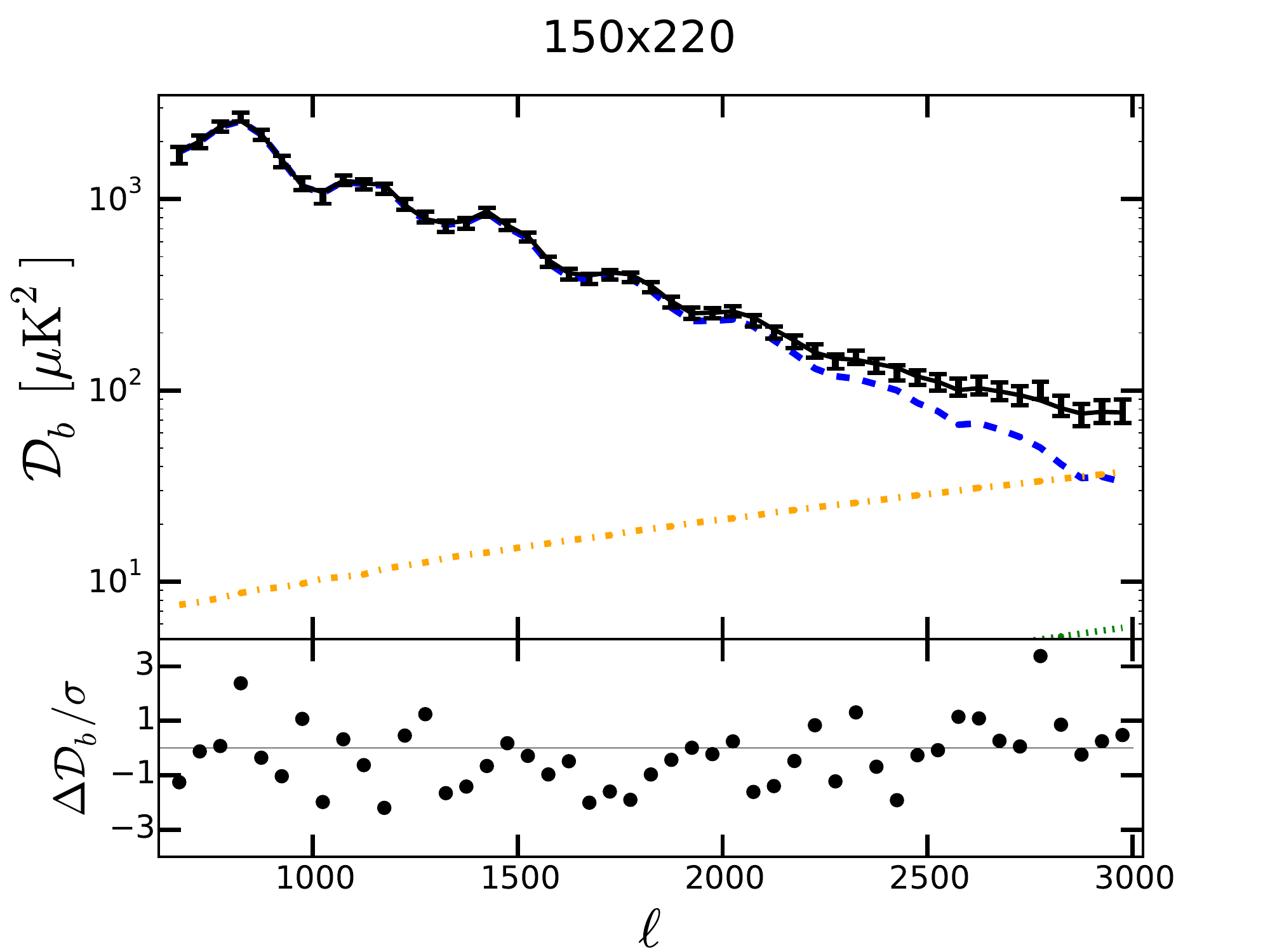}
\end{center}
\caption{
The six sets of single-frequency and cross-frequency bandpowers before foreground cleaning.
In each plot the top panel shows a single power spectrum; error bars
derived from the diagonal of the noise part of the bandpower covariance matrix
(multiplied by a factor of three to make them more visible); the best-fit 
CMB-only bandpowers, foreground model from G15, and extra unclustered source power
calculated using Equation~\ref{eqn:ptsrc}; and the sum of these components.
The bottom panel of each plot shows the residual 
of the data after subtracting these three components, divided by the square root 
of the diagonal elements of the noise covariance matrix.}
\label{fig:dlraw}
\end{figure*}

Figure~\ref{fig:dlraw} shows the six sets of power spectra before 
foreground cleaning. 
Error bars are derived from the diagonal of the noise 
part of the bandpower covariance matrix (no sample-variance, beam, calibration,
or foreground components are included). To make the error bars more visible, the $3 \sigma$
range is shown (rather than the typical $1 \sigma$). Overplotted are the best-fit
CMB-only bandpowers $\bar{D}_b$ from Equation~\ref{eqn:dlbest}, the best-fit
foreground model based on G15, the best-fit extra unclustered source power 
calculated using Equation~\ref{eqn:ptsrc} (see 
Section~\ref{sec:foregroundTreatment} for details), and the sum of these three
components. The bottom panel of each plot in Figure~\ref{fig:dlraw} shows the residual 
of the data after subtracting these three components, divided by the square root 
of the diagonal elements of the noise covariance matrix. 

In the limit that noise is the only contribution to the bandpower covariance, 
we expect the plotted noise-variance-scaled residuals
to be centered at zero with unit scatter and little correlation between points.
Many of the non-noise sources of variance in the data are strongly correlated
between bins, however, and contributions from these sources of variance
will be visible as long-wavelength structure in the residuals, especially in the 
lowest-noise frequency combinations.
A calibration factor has been applied to the data (except \oneone) 
to remove the overall calibration-related residual; these factors are well within 
the calibration uncertainty used to create the calibration contribution to the 
covariance matrix. 
By eye, there are no obvious discrepancies between the 
data and the sum of the three components plotted. There is some discernible 
long-wavelength structure in some of the residuals (for example a negative drift in
\oneone).
%; these are consistent with the correlation structure of some of the non-noise
%components of the covariance matrix (see Section~\ref{sec:cov} and Figure~\ref{fig:errors}).
If these long-wavelength residuals are consistent with, for example, 
our measured beam uncertainty or the uncertainty in the foreground model that has been subtracted, 
then they will be properly down-weighted in the \chisq\ calculation (Equation~\ref{eqn:chisq}) and will not 
result in an elevated \chisq\ value; if they arise from some other, unmodeled source, 
they will potentially result in an elevated \chisq\ value and low PTE.

\begin{figure*}
\begin{center}
    \includegraphics[width=0.98\textwidth]{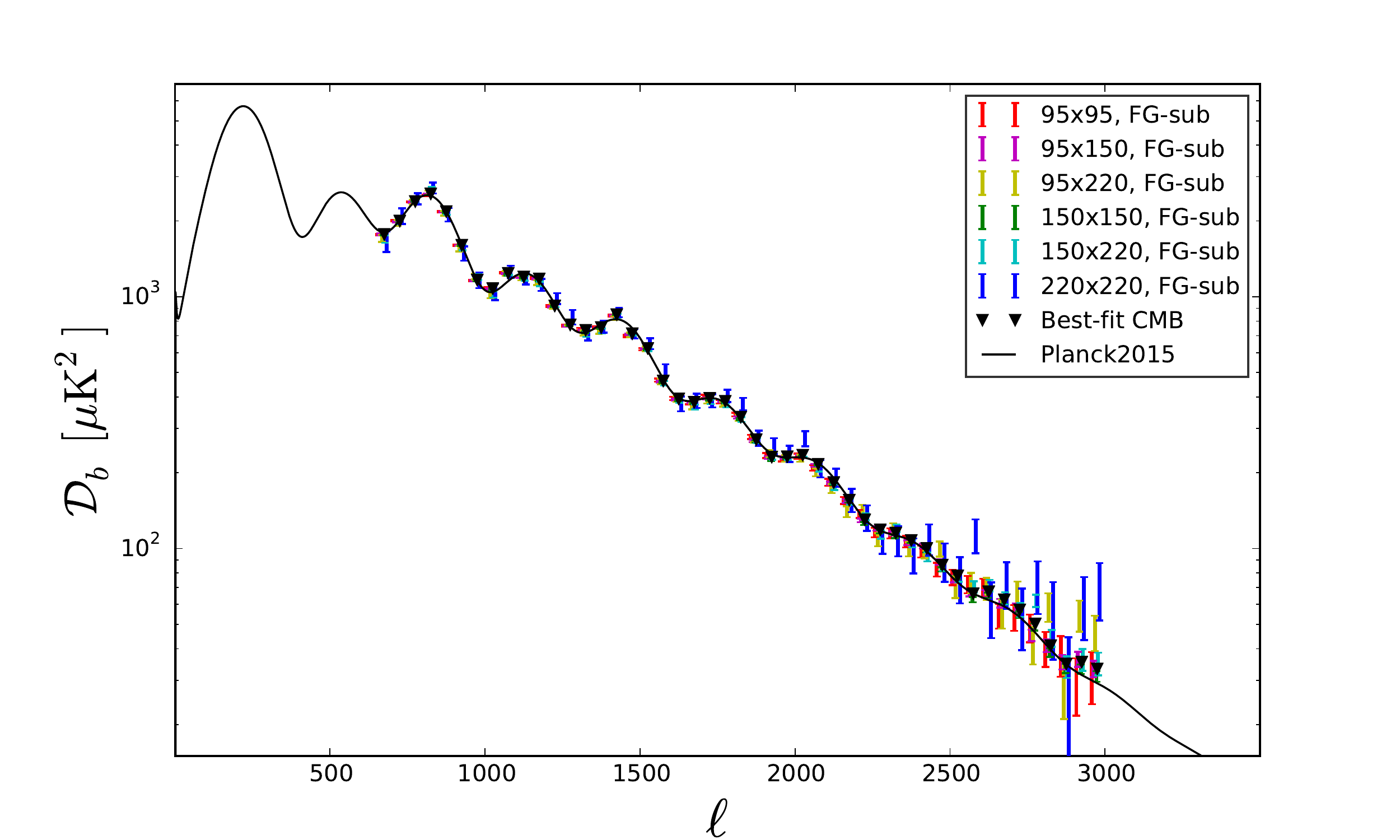}
\end{center}
\caption{
  The six individual foreground-subtracted single-frequency and cross-frequency spectra with the best-fit combined spectrum and the \planck\ 2015 
  {\tt TT+lowTEB+lensing} best-fit model overplotted.}
\label{fig:dl_auto}
\end{figure*}

The six foreground-cleaned power spectra, and the best-fit bandpowers $\bar{D}_b$,
are shown in Figure~\ref{fig:dl_auto}. 
%We show the bandpowers using the default method 
%of subtracting point-source power between the G15 source cut and the cut used here
%(see Section~\ref{sec:foregroundTreatment} for details).
The predicted temperature power spectrum from the 
best-fit \planck\ 2015 {\tt TT+lowTEB+lensing} cosmological model is overplotted as a guide, but we 
emphasize that neither this nor any other cosmological model was used in the fitting procedure.
As in Figure~\ref{fig:dlraw}, a calibration factor has been applied to the data for all spectra 
except \oneone. Again, the individual spectra look consistent by eye with the best-fit combined spectrum.

To make this statement more quantitatively, we compute the \chisq\ and associated PTE for
the model that the six foreground-cleaned power spectra consist of a single
common sky signal (the best-fit combined spectrum) and noise described by the bandpower
covariance matrix. The total number of points in the six spectra is 282 
(47 $\ell$-space bins per spectrum), and there are 47 free parameters in the fit (one value per bin
in the combined spectrum), so we have 235 remaining degrees of 
freedom. The \chisq, reduced \chisq, and PTE for the fit are:
\begin{eqnarray}
\chisq = \chisqfull \\      
\nonumber \chisq / \mathrm{dof} = \chisqredfull  \\
\nonumber \mathrm{PTE} = \ptefull
\end{eqnarray}
%
%If we use the secondary method of subtracting point-source power between the G15 cut
%and the cut used here, we obtain a \chisq\ of 229.6 and a PTE of 0.59.
Thus we find that the six power spectra are consistent with our simple
model, and we find no evidence of unmodeled systematics in the three-band SPT-SZ dataset.

We emphasize that, given the signal-to-noise on the power spectra in all six frequency combinations,
and given the removal of the ``protection'' of the large sample variance contribution to the total
covariance, this is a very stringent test of the consistency among the three SPT frequencies. 
The raw signal-to-noise in the six cross-spectra range from 10 to over 150 per bin, with a 
quadrature-summed signal-to-noise of over 1000. 
To give a sense of the level of unmodeled systematics
the test in this work is sensitive to, 
we have determined empirically that, for 
example, an unmodeled foreground contribution to the 150 GHz data at the level of a 
few $\mu$K$^2$ at $\ell \sim 3000$ would cause an unacceptable PTE, as would an unmodeled
instrumental systematic in one of the bands at the tens of $\mu$K$^2$ level at $\ell \sim 1000$.

We have also calculated the analogue to \chisq\ and PTE for each of the six individual frequency combinations
against the best-fit model. The interpretation of these quantities is not entirely straightforward, 
and we discuss these values and their interpretation in detail in Appendix~\ref{sec:appa}.
The general conclusion from the Appendix is that the distribution of \chisq\ values among the
six individual spectra is also consistent with expectations.

%%%%%%%%%%%%%%%%%%%%%%%%%%%%%%%%%%%%
% CONCLUSION
%%%%%%%%%%%%%%%%%%%%%%%%%%%%%%%%%%%%
\section{Conclusion}
\label{sec:conclusion}

We have conducted a consistency test of three-band data from 
the 2500-square-degree SPT-SZ survey. Using 95, 150, and 220~GHz maps---effectively the same data used in
previous SPT-SZ power spectrum analyses (including S13 and G15)---we have computed the 
six single-frequency and cross-frequency power-spectra among these three bands over the multipole range
$650 < \ell < 3000$. We have subtracted a model of foreground power from the spectra. 
The foreground model is based on the best-fit foreground
model in G15 with power added to account for sources that were masked in G15 but not in this analysis. Using a
bandpower covariance matrix with contributions from noise, uncertainties in the 
subtracted foreground model, and beam and calibration uncertainties, we conducted a
linear least-squares fit of the six spectra to a model in which each spectrum
consists solely of a common sky signal (assumed to be CMB anisotropy) and noise 
described by the covariance matrix. The reduced \chisq\ of that fit is \chisqredfull\ for 235 degrees
of freedom, for a PTE of \ptefull. We conclude from this result that there is no evidence of
unmodeled frequency-dependent systematic error in the three-band SPT-SZ data over this multipole range. 
Together with the null results from the comparison of the 150~GHz SPT-SZ data set 
to \planck\ data in two recent publications \cite{hou18,aylor17}, this result bolsters the 
conclusion that unmodeled systematics in SPT-SZ data are unlikely to be the cause of
any differences in cosmological parameters derived from SPT and \planck.

\begin{acknowledgments}
The South Pole Telescope program is supported by the National Science Foundation through grant PLR-1248097.
Partial support is also provided by the NSF Physics Frontier Center grant PHY-0114422 to the Kavli Institute of Cosmological Physics at the University of Chicago, the Kavli Foundation, and the Gordon and Betty Moore Foundation through grant GBMF\#947 to the University of Chicago.  
This research used resources of the National Energy Research Scientific Computing Center, which is supported by the Office of Science of the U.S. Department of Energy under Contract No. DE-AC02-05CH11231, and the resources of the University of Chicago Computing Cooperative (UC3), supported in part by the Open Science Grid, NSF grant NSF PHY 1148698.
CR acknowledges support from an Australian Research Council's Future Fellowship (FT150100074). 
Some of the results in this paper have been derived using the HEALPix \cite{gorski05} package. 
We acknowledge the use of the Legacy Archive for Microwave Background Data Analysis (LAMBDA). 
Support for LAMBDA is provided by the NASA Office of Space Science.
\end{acknowledgments}

%{\it Facilities:}
%\facility{South Pole Telescope}

\bibliography{../../BIBTEX/spt}

%%%%%%%%%%%%%%%%%%%%%%%%%%%%%%%%%%%%
% Appendix
%%%%%%%%%%%%%%%%%%%%%%%%%%%%%%%%%%%%
\appendix

\section{Single-spectrum Squared Deviation}
\label{sec:appa}

In this appendix, we discuss the analogue to \chisq\ and PTE for the six single-frequency
or cross-frequency power spectra compared to the best-fit model. We discuss how we create 
expectation values for these \chisq\ analogues from simulations, we report these expectation values and
the values from the data, and we calculate and discuss the comparison PTEs.

\subsection{Expected and Measured Single-frequency Weighted, Squared Deviation}

As discussed in Section~\ref{sec:results}, we fit a model with 47 free parameters
to a data set with 282 semi-independent data points, and we expect the sum of the weighted, 
squared residuals of the fit to follow a \chisq\ distribution with $282-47=235$ degrees
of freedom. We can also calculate the weighted, squared residuals for any subset
of the data. It is of particular interest to calculate the residuals for each of the six sets
of 47 bandpowers from the individual single-frequency or cross-frequency power spectra
(\ninenine, \nineone, etc.). The \chisq\ value for the entire data set is (cf. Equation~\ref{eqn:chisq})
\begin{equation}
\chisq = (D_\alpha - M_\alpha)^T  \ W_{\alpha \beta} \ (D_\beta - M_\beta).
\end{equation}
If we want to preserve all of the information about correlations
among the individual spectra, we can create the residual vector
$r_\alpha = D_\alpha - M_\alpha$ and its weighted counterpart 
$r^w_\alpha = W_{\alpha \beta} \ (D_\beta - M_\beta)$. The full \chisq\
is the sum of the product of $r$ and $r^w$, but we can also extract sums of subsets
of that product as the analogue to \chisq\ for the individual spectra. 
One might naively expect each set of these weighted, squared residuals
to follow a \chisq\ distribution with $235/6 \simeq 39$ degrees of freedom. This is, however,
only the case in the scenario in which the data points in the six sets of spectra are statistically
independent, and each individual power spectrum has equal weights in the fit. The second
point is elaborated upon in the next section.

%Instead of simply comparing the squared deviation of each set of 47 points to the
%same number of degrees of freedom, 
We calculate expectation values for the weighted, squared
deviation for each individual spectrum from simulations, and we compare the
data to those expectations. 
The ``simulations'' we use for this purpose
are simply the sum of mock CMB bandpowers and Gaussian realizations of 
the $282 \times 282$-element covariance matrix $C_{\alpha \beta}$ used in the full fit. For each 
realization, we create a 47-element mock CMB bandpower vector (using a Gaussian realization
of a theory power spectrum, binned using the same binning we use in the data) 
and add six copies of it to the 282-element realization of the covariance matrix. We perform the 
fit described in Section~\ref{sec:consistall} on each of these sets of mock bandpowers, and we calculate
the total squared deviation for each of the six (47-element) individual spectra:
\begin{equation}
\mathrm{WSD}_i = \sum_{\alpha=47i}^{47i+46} r_\alpha r^w_\alpha.
\end{equation}
We then perform the same calculation on the real data and compare the individual 
squared deviation values to the expected values for that spectrum. Table~\ref{tab:sdsingle}
shows for each individual spectrum the mean value of squared deviation from simulations, 
the value in the real data, and the fraction of simulations with higher squared deviation 
than the real data.

Because we already know from the main result of the paper that the ensemble data is 
consistent with expectations, we do not need to examine the overall distribution of these
squared deviation and PTE values but need only look for individual outliers. The
\oneone\ spectrum has the PTE farthest from 50\%, with only 3\% of simulations having
a higher squared deviation than the data. Using a simple trials-factor correction (from looking
at six individual PTEs, also known as the Bonferroni correction, e.g., \cite{dunn61}), we
would only flag this PTE value as significant if our threshold for single-PTE values was 
18\%. Thus we conclude the individual-spectrum deviations are also consistent with 
expectations.

\begin{table}[tbp]
\centering
\begin{tabular}{| l | l | l | l |}
\hline
spectrum & expected WSD & actual WSD & PTE \\
\hline
\ninenine & 39.58 & 27.50 & 0.83 \\
\nineone & 31.50 & 25.14 & 0.70 \\
\ninetwo & 44.97 & 54.47 & 0.19 \\
\oneone & 26.82 & 41.87 & 0.03 \\
\onetwo & 44.45 & 58.61 & 0.12 \\
\twotwo & 46.76 & 28.73 & 0.94 \\
\hline
\end{tabular}
\caption{\label{tab:sdsingle} Expected and actual total weighted, squared deviation values for the six individual spectra, 
and the fraction of simulations with higher values than the data for that spectrum.}
\end{table}

% 37.369947       24.534198       44.765555       23.434483       42.994347
%       46.663910

\subsection{Unequal Sharing of Degrees of Freedom}

The somewhat counter-intuitive result that the individual-spectrum weighted, squared 
deviations would not all be expected to be equal 
is easiest to understand in the simplified case of ``fitting'' $N$ 
uncorrelated variables to the model in which they have the same underlying value---i.e., 
taking the weighted average of $N$ numbers. For uncorrelated variables $y_i$ 
with uncertainty $\sigma_i$, the inverse-noise-weighted average is
\begin{eqnarray}
\bar{y} &=& \frac{\sum_i y_i/\sigma_i^2}{\sum_i 1/\sigma_i^2} \\
&\equiv& \frac{1}{\wtot} \sum_i \frac{y_i}{\sigma_i^2}.
\end{eqnarray}
Assuming the uncertainties are Gaussian-distributed, i.e., $P(y_i) = {\cal N}(\ytrue,\sigma_i)$,
where ${\cal N}(\mu,\sigma)$ is the normal distribution with mean $\mu$ and variance $\sigma^2$, the
probability distribution of $\bar{y}$, as expected, is 
\begin{eqnarray}
P(\bar{y}) &=& P\left (\frac{1}{\wtot} \sum_i \frac{y_i}{\sigma_i^2}\right ) \\
&=& 
\nonumber P\left (\frac{1}{\wtot} \frac{y_0}{\sigma_0^2}\right ) \circledast 
P\left (\frac{1}{\wtot} \frac{y_1}{\sigma_1^2}\right ) \circledast 
... \circledast 
P\left (\frac{1}{\wtot} \frac{y_N}{\sigma_N^2}\right ) \\
&=& 
\nonumber {\cal N}\left (\frac{\ytrue}{\wtot \sigma_0^2},\frac{1}{\wtot \sigma_0}\right ) \circledast 
{\cal N}\left (\frac{\ytrue}{\wtot \sigma_1^2},\frac{1}{\wtot \sigma_1}\right ) \circledast 
... \circledast 
{\cal N}\left (\frac{\ytrue}{\wtot \sigma_N^2},\frac{1}{\wtot \sigma_N}\right ) \\
&=& \nonumber {\cal N}\left (\frac{\ytrue}{\wtot \sum_i \sigma_i^2},\sqrt{\frac{1}{\wtot^2 \sum_i \sigma_i^2}}\right ) \\
&=& \nonumber {\cal N}\left (\ytrue,\sqrt{\frac{1}{\wtot}}\right ),
\end{eqnarray}
where $\circledast$ indicates convolution. The probability distribution of the residual
between any one $y_i$ and the best-fit $\bar{y}$ is:
\begin{eqnarray}
\label{eqn:presid}
P(y_i - \bar{y}) &=& P\left ( \left [1 - \frac{1}{\wtot \sigma_i^2} \right ] y_i - \frac{1}{\wtot} \sum_{j \neq i} \frac{y_j}{\sigma_j^2}\right ) \\
&=& 
\nonumber {\cal N}\left ( \left [1 - \frac{1}{\wtot \sigma_i^2} \right ] \ytrue, \left [1 - \frac{1}{\wtot \sigma_i^2} \right ] \sigma_i \right )
 \circledast 
{\cal N}\left ( - \frac{\ytrue}{\wtot \sigma_0^2}, \frac{1}{\wtot \sigma_0} \right ) \circledast ... \\
\nonumber &=& {\cal N} \left (0, \left [ \left (1 - \frac{1}{\wtot \sigma_i^2} \right ) \sigma_i^2 + \frac{1}{\wtot^2} \sum_{j \neq i} \frac{1}{\sigma_j^2} \right ]^{1/2} \right ) \\
\nonumber &\equiv& {\cal N}\left ( 0, \left [ \left (1 - f_1 \right )^2 \sigma_i^2 + \frac{f_2}{\wtot} \right ]^{1/2} \right ) \\
\nonumber &=& {\cal N}\left ( 0, \left [ f_2^2 \sigma_i^2  + f_2 \sigma_{\bar{y}}^2 \right ]^{1/2} \right ),
\end{eqnarray}
where $f_1 = 1/(\wtot \sigma_i^2)$ is the fraction of weight in sample $i$, 
$f_2 = 1/\wtot \sum_{j \neq i} 1/\sigma_j^2 = 1 - f_1$ is the fraction of weight in the
other samples, and $\sigma_{\bar{y}} = \sqrt{1/\wtot}$. 

If all the weights are equal, 
i.e., $\sigma_i = \sigma_0$ for all $i$, 
then $f_2 = (N-1)/N$, $\sigma_{\bar{y}} = \sigma_0/\sqrt{N}$, and
\begin{eqnarray}
P(y_i - \bar{y}) &=& {\cal N}\left ( 0, \left [ \left ( \frac{N-1}{N} \right )^2 \sigma_0^2 + \frac{N-1}{N^2} \sigma_0^2 \right ]^{1/2} \right ) \\
\nonumber &=& {\cal N}\left ( 0, \sqrt{1 - \frac{1}{N}} \ \sigma_0 \right ),
\end{eqnarray}
and every variable ``shares'' the loss of the one degree of freedom equally. But if the weights 
for the different $y_i$ are not identical, the sharing is not equal. In particular, in the limiting case 
of one variable dominating the weights, for that $y_i$, $f_2 \ll 1$, $\sigma_{\bar{y}} \simeq \sigma_i$, 
and 
\begin{equation}
P(y_i - \bar{y}) \simeq {\cal N}\left ( 0, \sqrt{f_2} \ \sigma_i \right),
\end{equation}
i.e., the variance of the residuals of this particular $y_i$ relative to the best-fit $y$ 
are heavily suppressed compared to variance of $y_i$ itself (because $y_i$ is dominating the fit).

Note that the total \chisq\ of the fit over all variables is not affected by the distribution of weights.
I.e., it can be shown from the variance of $y_i - \bar{y}$ in Equation~\ref{eqn:presid} that 
\begin{equation} 
\chisq = \sum_i \frac{(y_i - \bar{y})^2}{\sigma_i^2} = N - 1
\end{equation} 
regardless of the distribution of $\sigma_i$.

\end{document}